\documentclass[10pt,letterpaper]{article}
\usepackage[top=0.85in,left=2.75in,footskip=0.75in]{geometry}

\usepackage{amsmath,amssymb}

\usepackage{changepage}

\usepackage[utf8x]{inputenc}

\usepackage{textcomp,marvosym}

\usepackage{cite}

\usepackage{nameref,hyperref}
\hypersetup{
	colorlinks=true,
	linkcolor=blue,
	filecolor=magenta,      
	urlcolor=cyan,
}

\usepackage{microtype}
\DisableLigatures[f]{encoding = *, family = * }

\usepackage[table]{xcolor}

\usepackage{array}

\newcolumntype{+}{!{\vrule width 2pt}}

\newlength\savedwidth

\raggedright
\usepackage[aboveskip=1pt,labelfont=bf,labelsep=period,justification=raggedright,singlelinecheck=off]{caption}

\usepackage{authblk}
\usepackage{float}

\makeatletter
\renewcommand{\@biblabel}[1]{\quad#1.}
\makeatother

\usepackage{lastpage,fancyhdr,graphicx}
\usepackage{epstopdf}
\fancyheadoffset[L]{2.25in}
\fancyfootoffset[L]{2.25in}
\def \N{\,\mathcal{N}\,}
\newcommand\ee{{e}}
\newcommand\ii{{i}}
\def \tr{\,\mathrm{tr}\,}

\newcommand{\beginsupplement}{%
	\setcounter{table}{0}
	\setcounter{figure}{0}
	\setcounter{section}{0}
	\setcounter{equation}{0}
	\renewcommand{\thetable}{S\arabic{table}}%
	\renewcommand{\thefigure}{S\arabic{figure}}%
	\renewcommand{\theequation}{S\arabic{equation}}
	\renewcommand{\thesubsection}{S\arabic{subsection}}
	\renewcommand{\thesubsubsection}{\thesubsection.\arabic{subsubsection}}
}

\newenvironment{fullwidth}{%
	\begin{adjustwidth}{-2.25in}{0in}
	}{\end{adjustwidth}}

\RequirePackage[explicit]{titlesec}
\definecolor{MediumGrey}{HTML}{6D6E70}
\titleformat{\section}
{\color{MediumGrey}\Large\bfseries}
{\thesection}{}{#1}[]
\titleformat{\subsection}
{\large\bfseries}
{\thesubsection}{}{#1}[]
\titleformat{\subsubsection}
{\large}
{\thesubsubsection}{}{#1}[]    
\titleformat{\paragraph}
{\color{MediumGrey}\large}
{\theparagraph}{}{#1}[] 

\begin{document}
\vspace*{0.2in}
\begin{flushleft}
	{\LARGE\bfseries{Connecting empirical phenomena and theoretical models of biological coordination across scales}
	}
	\newline
	\\
	Mengsen Zhang\textsuperscript{1*},
	Christopher Beetle\textsuperscript{2},
	J. A. Scott Kelso\textsuperscript{1,3},
	Emmanuelle Tognoli\textsuperscript{1}
	\\
	\bigskip
	\textbf{1} Center for Complex Systems and Brain Sciences, Florida Atlantic University, Boca Raton, Florida, USA
	\\
	\textbf{2} Department of Physics, Florida Atlantic University, Boca Raton, Florida, USA
	\\
	\textbf{3} Intelligent System Research Centre, Ulster University, Derry$ \sim $Londonderry, Northern Ireland
	\\
	\bigskip
	
	* corresponding author: zhang@ccs.fau.edu
	
\end{flushleft}

\section*{Abstract}
  Coordination is ubiquitous in living systems. Existing theoretical models of coordination – from bacteria to brains – focus on either gross statistics in large-scale systems ($ N\rightarrow\infty $) or detailed dynamics in small-scale systems (mostly $ N=2 $). Both approaches have proceeded largely independent of each other. The present work bridges this gap with a theoretical model of biological coordination that captures key experimental observations of mid-scale social coordination at multiple levels of description. It also reconciles in a single formulation two well-studied models of large- and small-scale biological coordination (Kuramoto and extended Haken-Kelso-Bunz). The model adds second-order coupling (from extended Haken-Kelso-Bunz) to the Kuramoto model. We show that second-order coupling is indispensable for reproducing empirically observed phenomena and gives rise to a phase transition from mono- to multi-stable coordination across scales. This mono-to-multistable transition connects the emergence and growth of behavioral complexity in small and large systems.

\section{Introduction}

Coordination is central to living systems and biological complexity at large, where the whole can be more than and different from the sum of its parts. Rhythmic coordination \cite{Winfree1987,Kuramoto1984,Glass2001} is of particular interest for understanding the formation and change of spatiotemporal patterns in living systems, including e.g. slime mold \cite{Nakagaki2000,Kobayashi2006}, fireflies \cite{Buck1966,Mirollo1990}, social groups \cite{Neda2000nature,Richardson2007}, and the brain \cite{Grillner1985,Marder1996,Bressler2001,Kelso1995,Tognoli2014}. Theoretical descriptions of biological coordination phenomena are often in terms of coupled oscillators, whose behavior is constrained by their phase relations with each other \cite{Kuramoto1984,Winfree2001,Schoner1988,Rand1988,Kopell1988}. Existing studies of phase coordination  often focus on systems of either very few (small-scale, mostly $ N=2 $) \cite{Kelso1995,Jeka1993,Yokoyama2011}, or very many oscillators (large-scale, $ N\rightarrow\infty $) \cite{Acebr2005,Breakspear2010,Castellano2009}.  Here we inquire how the two might be connected and applied to midscale systems with neither too many nor too few components. The present work answers this question by modeling empirically observed coordinative behavior in midscale systems (N=8), based on data collected in an specially-designed human experiment \cite{Zhang2018}. The resultant model that captures all key experimental observations happens to also connect previous theories of small- and large-scale biological coordination in a single mathematical formulation.

But first, how are small- and large-scale models different? Small-scale models were usually developed to capture empirically observed coordination patterns, as in animal gaits \cite{Schoner1990}, bimanual movement coordination \cite{Kelso1984,Haken1985}, neuronal coordination \cite{Marder2001}, interpersonal coordination \cite{Tognoli2007,Schmidt2008}, human-animal coordination \cite{Lagarde2005} and human-machine coordination \cite{Kelso2009, Dumas2014}. They describe multiple stable coordination patterns (multistability) and the transitions between them (order-to-order transitions), e.g. from a trot to a gallop for a horse \cite{Hoyt1981}. In humans, dyadic coordination patterns like inphase and antiphase (synchronization, syncopation) were found across neural, sensorimotor, and social levels (see \cite{Kelso1995, Tognoli2014} for reviews), well captured by the extended Haken-Kelso-Bunz (HKB) model \cite{Haken1985,Kelso1990dw,Fuchs1996}. However, the extended HKB was restricted to describing coordination phenomena at $ N=2 $ (i.e. not directly applicable to higher-dimensional coordination phenomena). In contrast, large-scale models are concerned more about statistical features like the overall level of synchrony, disorder-to-order transitions, but not so much about patterns at finer levels. As a representative, the classical Kuramoto model \cite{Kuramoto1984} is applicable to describing a wide range of  large-scale coordination between, e.g., people \cite{Castellano2009,Neda2000PRE}, fish \cite{Leonard2012}, and neural processes \cite{Breakspear2010}, often studied analytically for its incoherence-to-coherence transition (at the statistical level, for $ N\rightarrow \infty$; see \cite{Strogatz2000,Acebron1998} for reviews). 

Although the extended HKB and the classical Kuramoto model emerged separately, they connect to each other by an interesting difference: the Kuramoto model with $ N=2 $ is \textit{almost} the extended HKB model except that the former lacks the term responsible for antiphase coordination in the latter (more accurately, the bistability of inphase and antiphase). Bistability of inphase and antiphase coordination, with associated order-to-order transitions and hysteresis, happens to be a key observation in small-scale human experiments \cite{Kelso1984,Kelso2012}. This begs the question of whether there is a fundamental difference between large-scale and small-scale coordination phenomena. Does the existence of antiphase, multistability, and order-to-order transitions depend on scale $ N $? With these questions in mind, we recently conducted a human experiment \cite{Zhang2018} at an intermediate scale ($ N=8 $), such that the system is large enough for studying its macro-level properties, yet small enough for examining patterns at finer levels, ideal for theories and empirical data to meet at multiple levels of description. In the following sections, we demonstrate how the marriage between the two models (not either one alone) is sufficient for capturing empirical observations at multiple levels of description and discuss its empirical and theoretical implications for biological coordination.

\section{Results}
\subsection{Human coordination at intermediate scales}

\begin{figure}[H]
	\centering
	\includegraphics[width=0.7\textwidth]{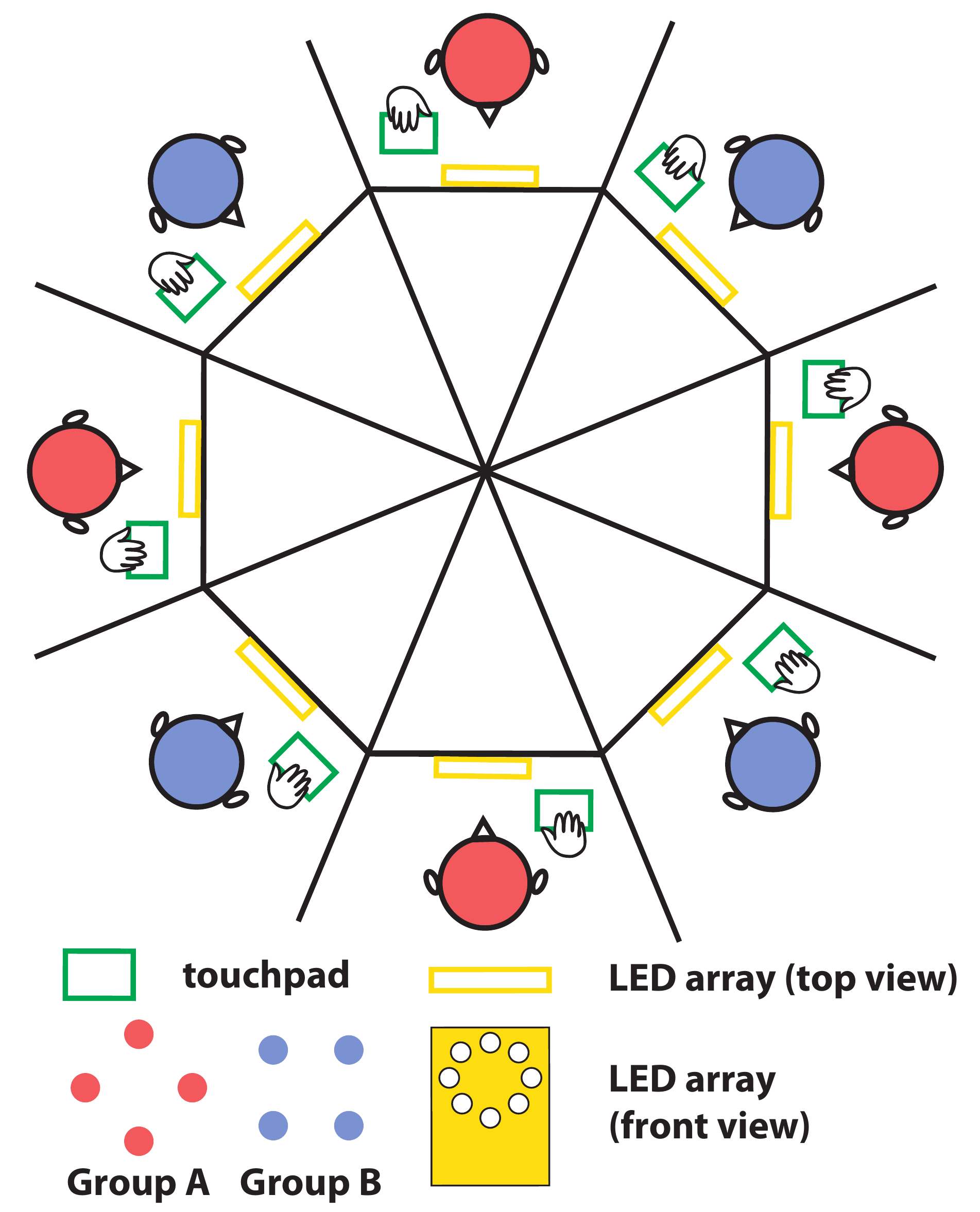}
	\caption{Experimental setup for multiagent coordination. In the Human Firefly experiment \cite{Zhang2018}, eight subjects interacted simultaneously with each other via a set of touch pads and LED arrays. In each trial, each subject was paced with a metronome prior to interaction. The metronome assignment split the ensemble of eight into two frequency groups of four (group A and B, colored red and blue respectively). The frequency difference $ \delta f $ between group A and B was systematically manipulated to induce different grouping behavior. See text for details. }
	\label{fig:setup}
\end{figure}

\begin{figure}[H]
	\begin{fullwidth}
		\includegraphics[width=\linewidth]{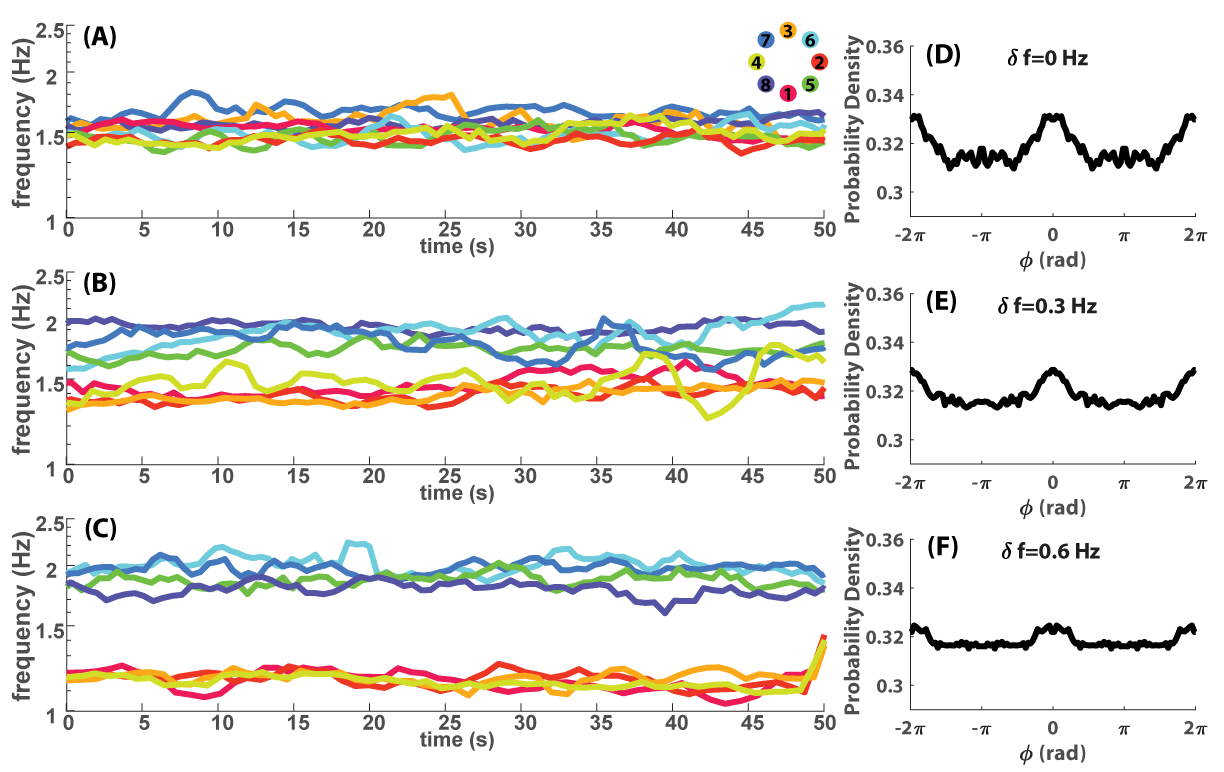}
		\caption{Examples of frequency dynamics and aggregated relative phase distributions for three diversity conditions. (A-C) shows instantaneous frequency (average over four cycles) from three trials with diversity $ \delta f=0,\,0.3,\,0.6 $ Hz respectively. Viewed from bottom to top, in (C), two frequency groups of four are apparent and isolated due to high intergroup difference (low-frequency group, warm colors, paced with metronome $ f_A=1.2 $ Hz; high-frequency group, cold colors, paced with metronome $ f_B=1.8 $ Hz). As the two groups get closer (B), more cross-talk occurred between them (note contacting trajectories especially after 30s). Finally when the intergroup difference is gone (A), one supergroup of eight formed. (D-F) show relative phase $ \phi $ distributions aggregated from all trials for $ \delta f=0,\,0.3,\,0.6 $ Hz respectively (histograms computed in $ [0,\pi) $, plotted in $ [-2\pi,2\pi] $ by symmetry and periodicity). When diversity is low (D), the distribution peaks near inphase ($ \phi=0 $) and antiphase ($ \phi=\pi $), separated by a trough near $ \pi/2 $, with antiphase weaker than inphase. The two peaks are diminished as $ \delta f $ increases (E,F), but the weaker one at antiphase becomes flat first (F).}\label{fig:3df_example_phi}
	\end{fullwidth}
\end{figure}

Before getting into the model, we briefly review the mid-scale experiment and key results \cite{Zhang2018}. In the experiment (dubbed the ``Human Firefly" experiment), ensembles of eight people ($ N=8$, total 120 subjects) spontaneously coordinated rhythmic movements in an all-to-all network (via 8 touchpads, and 8 ring-shaped arrays of 8 LEDs as in Figure~\ref{fig:setup}; see \nameref{section:methods} for details). To induce different grouping behavior, subjects were paced with different metronomes prior to interaction such that each ensemble was split into two frequency groups of equal size with intergroup difference $ \delta f=0$, $ 0.3 $, or $ 0.6 $ Hz (referred to as levels of ``diversity"), and were asked to maintain that frequency during interaction after the metronome was turned off. Subjects' actual frequencies from three example trials (Figure~\ref{fig:3df_example_phi}A-C) show intuitively the consequences of frequency manipulations: from (A) to (C) a supergroup of eight gradually split into two frequency groups of four as diversity increased from $ \delta f=0 $ Hz to $ 0.6 $ Hz. 

\begin{figure}[H]
	\begin{fullwidth}
		\centering
		\includegraphics[width=\linewidth]{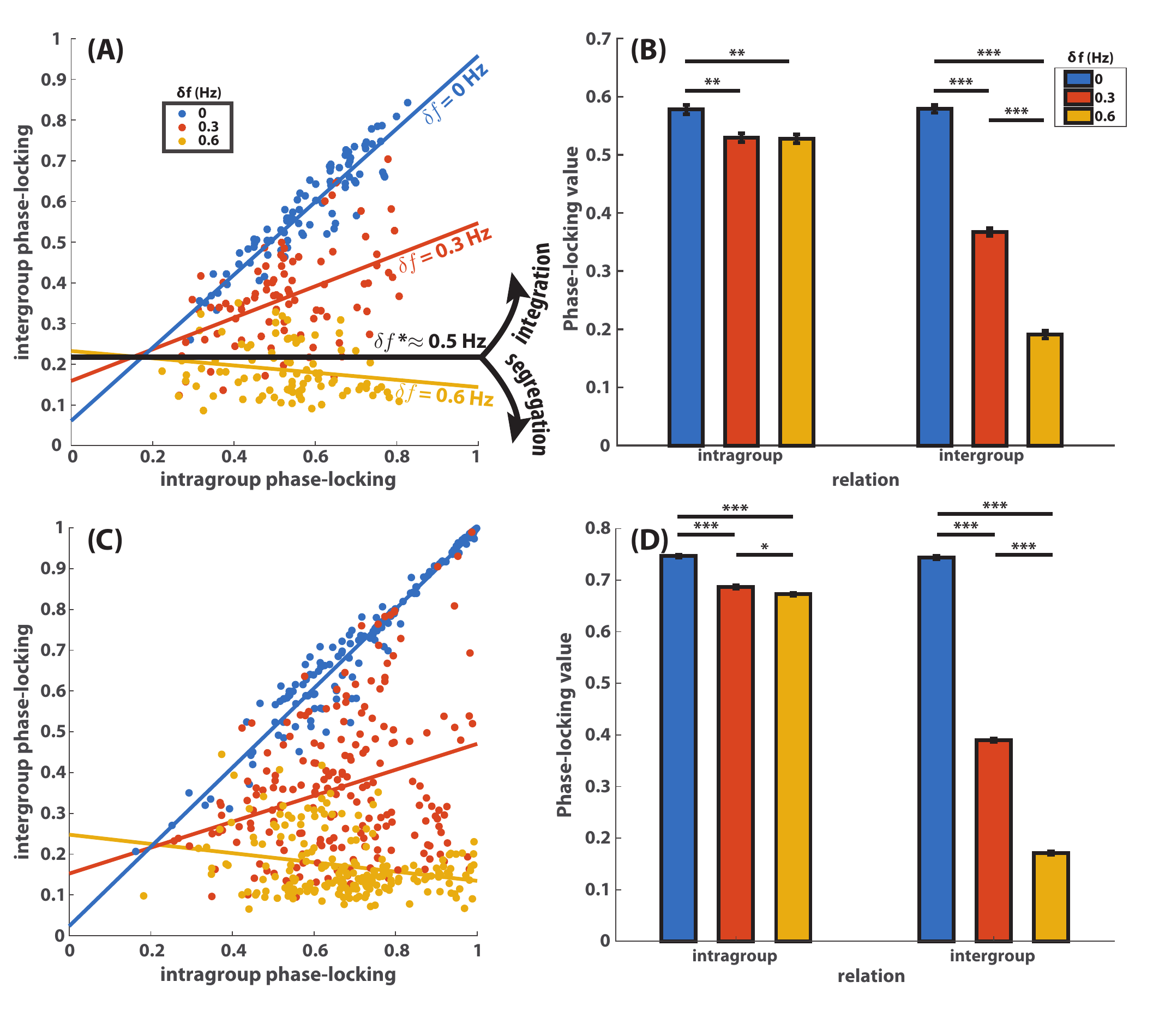}
		\caption{
			Intergroup relations and average inter/intragroup coordination. (A)  shows how intragroup coordination relates to intergroup coordination for different levels of diversity ($ \delta f $, color-coded) in the ``Human Firefly" experiment \cite{Zhang2018}. Each dot's x- and y-coordinate reflect the level of intragroup and intergroup coordination respectively (measured by phase-locking value) for a specific trial. Lines of corresponding colors are regression lines fitted for each diversity condition (slope $ \beta_1 $ indicates the level of integration between groups). With low and moderate diversity (blue and red), two frequency groups are integrated (positive slopes); and with high diversity (yellow), two frequency groups are segregated (negative slope). Black line (zero slope) indicates the empirically estimated critical diversity $ \delta f^* $, demarcating the regimes of intergroup integration and segregation. The exact same analyses were applied to the simulated data (200 trials per diversity condition) and the results are shown in (C), which highly resemble their counterparts in (A). (B) shows a break-down of the average level of dyadic coordination as a function of diversity (color) and whether the dyadic relation was intragroup (left) or intergroup (right). Intragroup coordination was reduced by the presence of intergroup diversity ($ \delta f\neq 0 $; left red, yellow bars shorter than left blue bar); intergroup coordination dropped rapidly with increasing $ \delta f $ (right three bars; error bars reflect standard errors). Results of the same analyses on simulated data are shown in (D), which again highly resemble those of the human data in (B). 
		}
		\label{fig:criticaldf_anova}
	\end{fullwidth}
\end{figure}

Key results involve multiple levels of description, in terms of intergroup, intragroup and interpersonal relations. The level of intergroup integration is defined as the relation between intragroup and intergroup coordination ($ \beta_1 $, slope of regression lines in Figure~\ref{fig:criticaldf_anova}A; see \nameref{section:methods}). Two frequency groups are integrated when diversity is low or moderate ($ \delta f=0,\,0.3 $Hz, blue and red lines, slope $ \beta_1>0 $) and segregated when diversity is high ($ \delta f=0.6 $ Hz, yellow line, slope $ \beta_1<0 $). A critical level of diversity demarcating the regime of intergroup integration and segregation was estimated to be $ \delta f^*=0.5 $ Hz. Within the frequency groups, coordination was also reduced by the presence of intergroup difference (Figure~\ref{fig:criticaldf_anova}B, left, red and yellow bars shorter than blue bar). At the interpersonal level, inphase and antiphase were preferred phase relations (inphase much stronger than antiphase; distributions in Figure~\ref{fig:3df_example_phi}D-F), especially when the diversity was very low (Figure~\ref{fig:3df_example_phi}D, peaks around $ \phi=0,\,\pi $, in $ radians $ throughout this paper), but both were weakened by increasing diversity (Figure~\ref{fig:3df_example_phi}EF; in episodes of strong coordination, antiphase is greatly amplified, and much more susceptible to diversity than inphase, see \cite{Zhang2018}). Notice that subjects did not lock into these phase relations but rather engaged and disengaged intermittently (two persons dwell at and escape from preferred phase relations recurrently, a sign of metastability; see Figure~\ref{fig:triad}A red trajectory for example), reflected also as ``kissing" and ``splitting" of frequency trajectories (e.g. in Figure~\ref{fig:3df_example_phi}B). In the following sections, we present a model that captures these key experimental observations at their respective levels of description. 

\subsection{A minimal experiment-based model of multiagent coordination}

Our model of coordination is based on a family of $N$ oscillators, each represented by a single phase angle $\varphi_i$.  We will show that a pair-wise phase coupling \cite{Kuramoto1984,Schoner1990,Haken1985} of the form 

\begin{equation}\label{eqn:phaseorder2}
\dot\varphi_i = \omega_i - \sum_{j = 1}^N a_{ij} \sin (\varphi_i - \varphi_j) - \sum_{j = 1}^N b_{ij} \sin 2 (\varphi_i - \varphi_j)
\end{equation}
suffices to model the key features of the experimental data identified above.  The left side of this equation is the time derivative of $\varphi_i$, while the constant $\omega_i > 0$ on the right is the natural (\textit{i.e.}, uncoupled) frequency of the $i^{\mathrm{th}}$ oscillator.  The coefficients $a_{ij} > 0$ and $b_{ij} > 0$ are parameters that govern the coupling.

The equations (\ref{eqn:phaseorder2}) include a number of well-studied models as special cases.  For instance, setting $\phi := \varphi_1 - \varphi_2$, $\delta\omega := \omega_1 - \omega_2$, $\tilde{a} := a_{12} + a_{21}$, and $2 \tilde{b} := b_{12} + b_{21}$ for $N = 2$, the difference of the two resulting equations (\ref{eqn:phaseorder2}) yields the relative phase equation 

\begin{equation}
\dot\phi = \delta\omega - \tilde{a} \sin \phi - 2 \tilde{b} \sin 2\phi \label{eqn:HKB}
\end{equation} 
of the extended HKB model \cite{Kelso1990dw}. The HKB model \cite{Haken1985} was originally designed to describe the dynamics of human bimanual coordination, corresponding to equation {(\ref{eqn:HKB})} with $ \delta\omega=0$ (i.e. describing the coordination between two identical components). The extended HKB introduces the symmetry breaking term $ \delta\omega $ to capture empirically observed coordinative behavior between asymmetric as well as symmetric components (i.e. the HKB model is included in the extended HKB model, which is further included in equations \ref{eqn:phaseorder2}). It has since been shown to apply to a broad variety of dyadic coordination phenomena in living systems, e.g. \cite{Jeka1993,Schmidt1994,Kelso1995,Kelso2012,Tognoli2014}.  Equations (\ref{eqn:phaseorder2}) can be considered a generalization of the extended HKB model from 2 to $N$ oscillators.  It is remarkable that such a direct generalization can reproduce key features of the collective rhythmic coordination in \emph{ensembles} of human subjects at \emph{multiple} levels of description.

Another well-studied special case of equations (\ref{eqn:phaseorder2}) is the Kuramoto model \cite{Kuramoto1984}, which has $b_{ij} = 0$ (and typically $a_{ij} = a$, independent of $i$ and $j$).  We will see below, however, that the Kuramoto model cannot exhibit at least one feature of the experimental data.  Namely, the data show a secondary peak in the pairwise relative phase of experimental subjects at antiphase (see Figure~\ref{fig:3df_example_phi}D-F above).  Simulations using the Kuramoto model do not reproduce this effect, while simulations of equations (\ref{eqn:phaseorder2}) model do (compare Figure~\ref{fig:modeldyn_phi} D-F and G-I below).  We give additional analytical support for this point by studying relevant fixed points of both models in the \nameref{SI} (Section \nameref{section:CB_multistability}).

\subsection{Weak coupling captures human behavior}
Given the spatially symmetric setup of the ``Human Firefly" experiment (all-to-all network, visual presentation at equal distance to fixation point), it is  reasonable to further simplify equations (\ref{eqn:phaseorder2}) by letting $ a_{ij}=a $ and $ b_{ij}=b $ ($ a,\,b>0 $),

\begin{align}
\dot{\varphi}_{i}=\omega_i-a\sum_{j=1}^{N}\sin \phi_{ij}-b\sum_{j=1}^{N}\sin 2\phi_{ij}\label{eqn:uniformNHKB}
\end{align}
where $ \phi_{ij}=\varphi_{i}-\varphi_{j} $ is the relative phase between oscillators $ i $ and $ j $ (henceforth we use the notation $ \phi_{ij} $ instead of the subtraction, since relative phase is the crucial variable for coordination \cite{Kelso1984,Haken1985}).

At the level of intergroup relations, model behavior (Figure~\ref{fig:criticaldf_anova}C; under weak coupling $ a=b=0.105 $; see \nameref{section:parspace} in \nameref{SI} on parameter choices) successfully captures human behavior (Figure~\ref{fig:criticaldf_anova}A) at all levels of diversity. Similar to the human experiment, low diversity ($ \delta f=0 $ Hz) results in a high level of integration in the model (blue line in Figure~\ref{fig:criticaldf_anova}C slope close to 1; $ \beta_1 =0.972$, $ t(199)=66.6 $, $ p<0.001 $); high diversity ($ \delta f=0.6 $ Hz) comes with segregation (yellow line slope negative; $ \beta_1=-0.113$, $ t(199)=-3.56 $, $ p<0.001 $); and in between, moderate diversity ($ \delta f=0.3 $ Hz) is associated with partial integration (red line positive slope far less than 1; $ \beta_1=0.318 $, $ t(199)=4.23 $, $ p<0.001 $). Here we did not estimate the critical diversity $ \delta f^* $ the same way as for the human data (by linear interpolation), since we found theoretically that the level of integration depends nonlinearly on diversity $ \delta f $, and as a result the theoretical $ \delta f^* $ is $ 0.4 $ Hz (see Figure~\ref{fig:parspace_NHKB}D). This prediction can be tested in future experiments by making finer divisions between $ \delta f=0.3 $ and $ 0.6 $ Hz. 

\begin{figure}[H]
	\begin{fullwidth}
		\includegraphics[width=\linewidth]{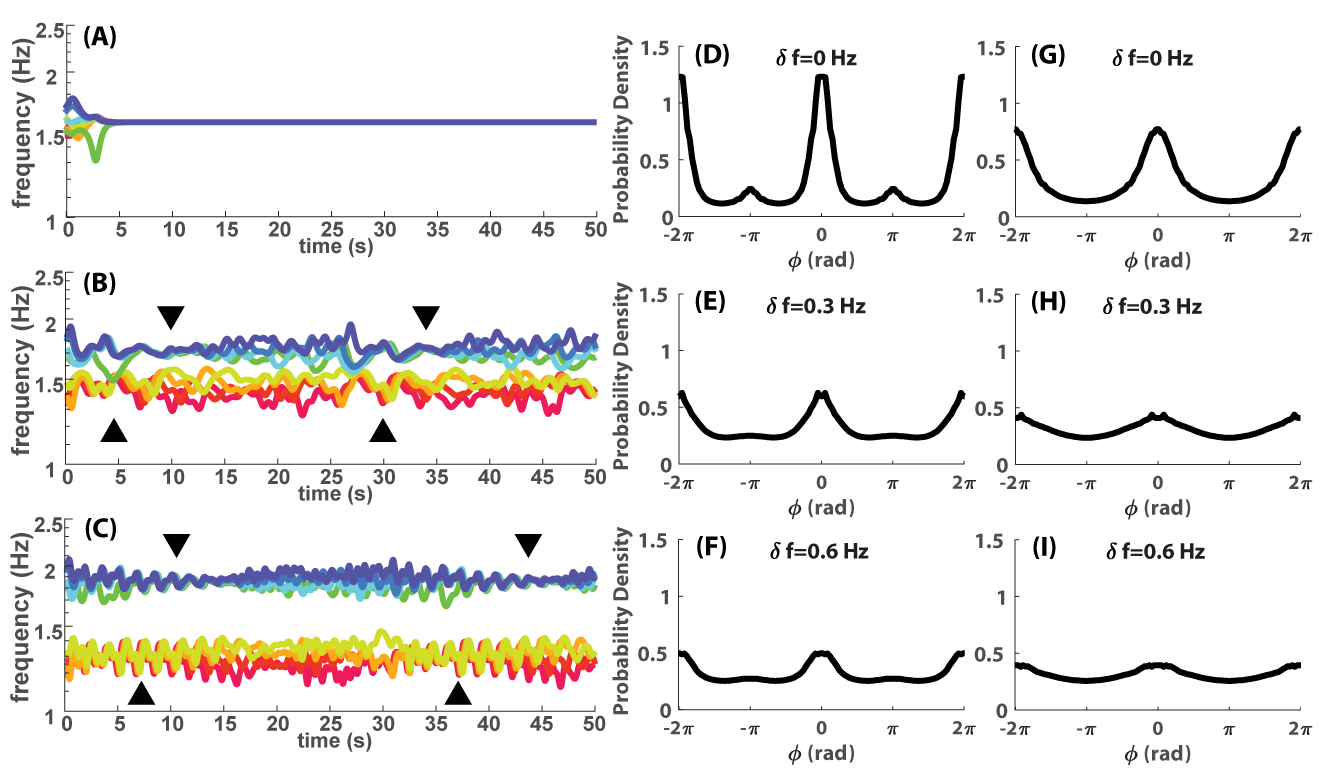}
		\caption{Examples of frequency dynamics and aggregated relative phase distributions for simulated data. (A-C) show frequency dynamics of three simulated trials ($ a=b=0.105 $) with the same initial phases and intragroup frequency dispersion but different intergroup difference i.e. $ \delta f=0,\, 0.3,\, 0.6$ Hz respectively. When intergroup differences are introduced (BC), not only is intergroup interaction altered but intragroup coordination also loses stability and becomes metastable (within-group trajectories converge at black triangles and diverge afterwards). The time scale of metastable coordination also changes with $ \delta f $, i.e. the inter-convergence interval was shorter for (B) than (C). (D-F) show relative phase distributions, aggregated over 200 trials ($ a=b=0.105 $) for each diversity condition ($\delta f=0,\,0.3,\,0.6 $ respectively). At low diversity (D), there is a strong inphase peak and a weak antiphase peak, separated by a trough near $ \pi/2 $. Both peaks are diminished by increasing diversity (EF). These features are qualitatively the same as the human experiment. (G-I) show the same distributions as (D-F) but for $ a=0.154 $ and $ b=0 $ (i.e. the classical Kuramoto model). There is a single peak in each distribution at inphase $ \phi=0 $, and a trough at antiphase $ \phi=\pi $.  }\label{fig:modeldyn_phi}
	\end{fullwidth}
\end{figure}

In the human experiment, not only did we uncover the effect of diversity on intergroup relations, but also, non-trivially, on intragroup coordination (outside affects within, a sign of complexity). Statistically, this is shown in Figure~\ref{fig:criticaldf_anova}B (three bars on the left): with the presence of intergroup difference ($ \delta f >0 $), intragroup coordination was reduced (red, yellow bars significantly shorter than blue bar). This is well captured by the model as shown in Figure~\ref{fig:criticaldf_anova}D (2-way ANOVA interaction effect, $ F(2,19194)=3416 $, $ p<0.001 $; the simulated data also capture the rapid decline of intergroup coordination with increasing $ \delta f $ in human data, shown in Figure~\ref{fig:criticaldf_anova}BD, right). To see what this means dynamically, three simulated trials are shown in Figure~\ref{fig:modeldyn_phi}A-C as examples (same initial conditions and intragroup frequency dispersion). The phase-locked state within groups (when $ \delta f=0 $ Hz; Figure~\ref{fig:modeldyn_phi}A) is lost and replaced by metastable coordination (intermittent convergence, marked by black triangles in Figure~\ref{fig:modeldyn_phi}BC) as soon as two groups begin to differentiate from each other ($ \delta f=0.3,0.6 $ Hz). In fact, the statistical result (Figure~\ref{fig:criticaldf_anova}B, left) reflects how two groups collaboratively increased each other's intragroup coordination (see \nameref{section:modulardynamics} for baseline dynamics when intergroup coupling is removed, and \nameref{section:noingroupvarfreq} in \nameref{SI} for statistics when intragroup variability is removed). Comparing Figure~\ref{fig:modeldyn_phi}B with C, we see the time scale of metastable convergence is also altered by intergroup difference $ \delta f $ (longer inter-convergence interval for C) - intergroup difference changes not only the overall \textit{level} of coordination within groups, but also the \textit{patterns} of coordination.

At interpersonal level, human subjects tended to coordinate with each other around inphase and antiphase, especially when the diversity is low ($ \delta f=0 $ Hz; Figure~\ref{fig:3df_example_phi}D, peaks around $ \phi=0,\,\pi $ separated by a trough near $ \phi=\pi/2 $); and the preference for inphase and antiphase both diminishes as diversity increases ($ \delta f=0.3,\, 0.6 $, Figure~\ref{fig:3df_example_phi}EF). Both aspects are well reproduced in simulations of the model (Figure~\ref{fig:modeldyn_phi}D-F). Note that these model-based distributions are overall less dispersed than the more variable human-produced distributions (Figure~\ref{fig:3df_example_phi}D-F), likely due to the deterministic nature of the model (i.e. no stochastic terms; as shown in the present work, a deterministic model is sufficient for capturing key statiscal and dynamic features in the human experiment, thus preferred for simplicity).

\subsection{The necessity of second-order coupling}

Equation (\ref{eqn:uniformNHKB}) becomes the classical Kuramoto model \cite{Kuramoto1984} when $ b=0 $. We follow the same analyses as in the previous section but now for $ a=0.154 $ and $ b=0 $ (see \nameref{section:parspace_kura} in \nameref{SI} on parameter choices). The relationship between intragroup and intergroup coordination (Figure~\ref{fig:kura_regression}A; $ \beta_1(0Hz)=0.974 $, $ t(199)=53.2 $, $ p<0.001 $; $ \beta_1(0.3Hz)=0.292 $, $ t(199)=4.52 $, $ p<0.001 $; $ \beta_1(0.6Hz)=-0.011 $, $ t(199)=-0.41 $, $ p>0.05 $) resembles the case of $ b\neq 0 $ ($ a=b=0.105 $, Figure~\ref{fig:criticaldf_anova}C. A difference remains that for $ b=0 $, $ \beta_1(0.6Hz) $ is not significantly less than zero ($ p=0.68 $; Figure~\ref{fig:kura_regression}A yellow). The average level of intragroup and intergroup coordination also varies with diversity in the same way as the case of $ b \neq 0 $ (Figure~\ref{fig:kura_regression}b for $ b=0 $, interaction effect $ F(2,19194)=3737 $, $ p<0.001 $, compared to Figure~\ref{fig:criticaldf_anova}D for $ b\neq 0 $). In short, group-level statistical features can be mostly preserved without second order coupling (i.e. $ b=0 $). 

However, this is no longer the case when it comes to interpersonal relations. The distributions of dyadic relative phases are shown in Figure~\ref{fig:modeldyn_phi}G-I. Without second order coupling, the model does not show a preference for antiphase in any of the three diversity conditions, thereby missing an important feature of human social coordination (for additional comparisons between human and model behavior, see Section~\nameref{section:stats_antiphase} in \nameref{SI}). Analytically, we find that the coupling ratio $ \kappa=2b/a $ determines whether antiphase is preferred (for the simple case of identical oscillators, in \nameref{section:CB_multistability} in \nameref{SI}). A critical coupling ratio $ \kappa_c=1 $ demarcates the regimes of monostability (only all-inphase is stable for $ \kappa<1 $) and multistability (any combination of inphase and antiphase is stable for $ \kappa>1 $). This critical ratio (for equation \ref{eqn:uniformNHKB}) is identical to the critical coupling of the HKB model \cite{Haken1985}, where the transition between monostability (inphase) and multistability (inphase and antiphase) occurs (equation \ref{eqn:HKB}, parameters in the two equations map to each other by $ a=\tilde{a}/2 $ and $ b=\tilde{b}$). This shows how equation (\ref{eqn:uniformNHKB}) is a natural N-dimensional generalization of the extended HKB model, in terms of multistability and order-to-order transitions.

\subsection{The effect of non-uniform coupling}

\begin{figure}[H]
	\begin{fullwidth}
		\centering
		\includegraphics[width=\linewidth]{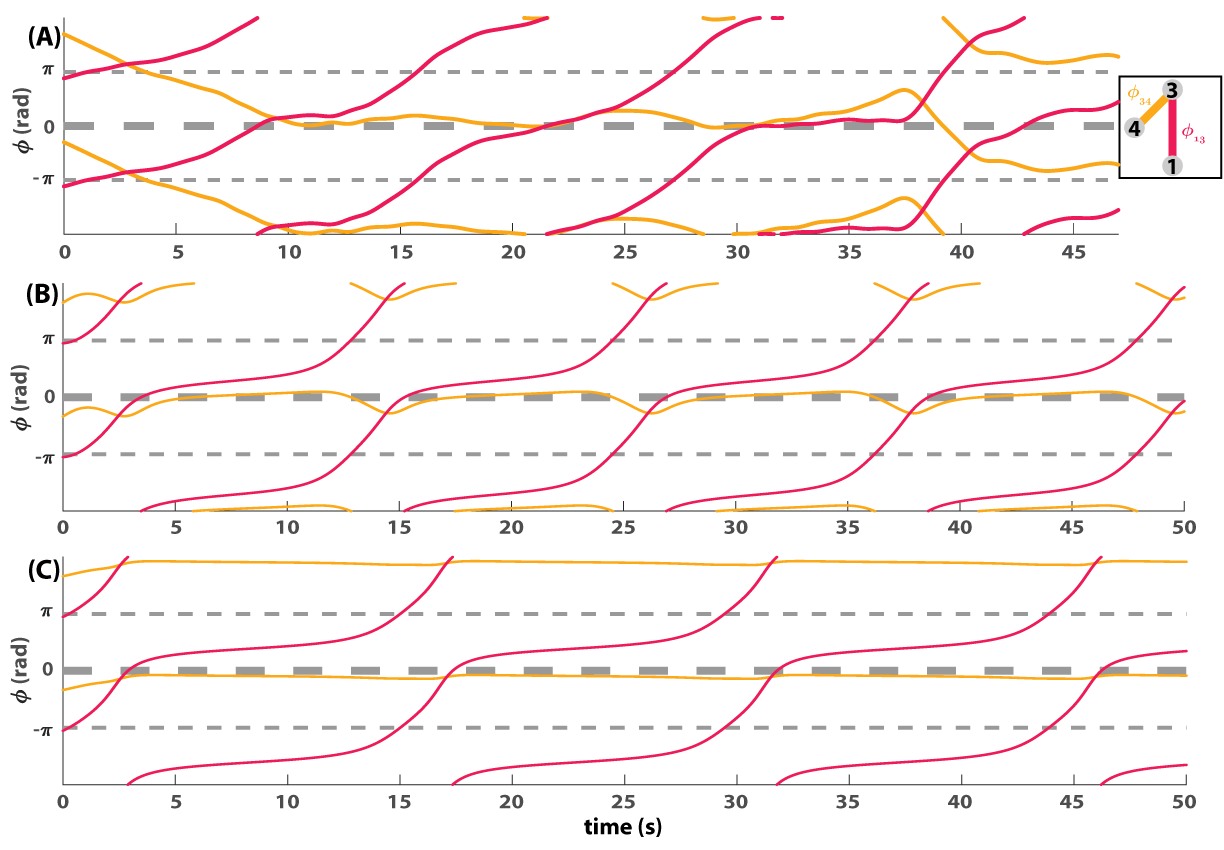}
		\caption{The effect of non-uniform coupling strength on coordination dynamics. (A) shows the evolution of the relationship between three persons (agent 1, 3, 4, spatially situated as in legend) in terms of two relative phases ($ \phi_{13} $, $ \phi_{34} $). $ \phi_{34} $ (yellow) persisted at inphase for a long time (10-37s trajectory flattened near $ \phi=0 $) before switching to antiphase (40s). $ \phi_{13} $ (red) dwelled at inphase intermittently (flattening of trajectory around 10, 20, and 35s). Three bumps appeared in $ \phi_{34} $ during its long dwell at inphase (near 15, 25, 37s), which followed the dwells in $ \phi_{13} $, indicating a possible influence of $ \phi_{13} $ on $ \phi_{34} $. (B,C) show two simulated trials with identical initial conditions and natural frequencies, estimated from the human data. In (B), agent 3 is more ``social" than agent 4 ($ a_3>a_4 $). More precisely, agent 3 has a much stronger coupling ($ a_3=1 $) than all others ($ a_1=a_4=b_1=b_3=b_4=0.105 $, as in previous sections). The recurring bumps in $ \phi_{34} $ are nicely reproduced. In (C), agent 3 and 4 are equally ``social" ($ a_3=a_4=0.5525 $, keeping the same average as in (B)). $ \phi_{34} $ is virtually flat throughout the trial.}
		\label{fig:triad}
	\end{fullwidth}
\end{figure}

So far, our model has captured very well experimental observations with the simple assumption of uniform coupling. However, loosening this assumption is necessary for understanding detailed dynamics. Here is an example from \cite{Zhang2018} (Figure~\ref{fig:triad}A), where coordination among three agents (1, 3, and 4, labels of locations on LED arrays) is visualized as the dynamics of two relative phases ($ \phi_{13} $ red, $ \phi_{34} $ yellow). Agents 3 and 4 coordinated inphase persistently (10-40s yellow trajectory flat at $ \phi_{34}\approx 0 $), while agents 3 and 1 coordinated intermittently every time they passed by inphase (red trajectory $ \phi_{13} $ becames flat, i.e. dwells, near inphase around 10, 20 and 35s). Curiously, every dwell in $ \phi_{13} $ (red) was accompanied by a little bump in $ \phi_{34} $, suggesting $ \phi_{34} $ was periodically influenced by $ \phi_{13} $. In the framework of our model, we can approximate the dynamics of $ \phi_{34} $ from equation (\ref{eqn:phaseorder2}) by assuming $ \phi_{34}=0 $ (thus $ \phi_{13}=\phi_{14} $),

\begin{align}
\dot{\phi}_{34}&=f(\phi_{34})+\underbrace{(a_{31}-a_{41})\sin \phi_{13}+(b_{31}-b_{41})\sin 2\phi_{13}}_{=:K(\phi_{13})}\label{eqn:phi34model}
\end{align}
where $ f(\phi_{34}) $ is the influence of $ \phi_{34} $ on itself, $ K(\phi_{13})$ the influence of $ \phi_{13} $ on $ \phi_{34} $. From $ K(\phi_{13})$ we see that $ \phi_{13} $ has no influence on $ \phi_{34} $ if the coupling is completely uniform (i.e. $ K(\phi_{13})\equiv 0 $ if $ a_{31}=a_{41} $ and $ b_{31}=b_{41} $), making it impossible to capture the empirical observation (red relation influencing yellow relation, Figure~\ref{fig:triad}A). To break the symmetry between agent 3 and 4, we ``upgrade" equation (\ref{eqn:uniformNHKB}) to the system

\begin{align}
\dot{\varphi}_{i}=\omega_i-a_i\sum_{j=1}^{N}\sin \phi_{ij}-b_i\sum_{j=1}^{N}\sin 2\phi_{ij} \label{eqn:NHKB_asymcouple}
\end{align}
where each oscillator can have its own coupling style (oscillator specific coupling strength $ a_i $ and $ b_i $). In the present case, we are interested in what happens when $ a_3\neq a_4 $ for $ i\in\{1,3,4\} $. Two simulated trials are shown in Figure~\ref{fig:triad}B and C with non-uniform vs. uniform coupling (same initial conditions and natural frequencies across trials, estimated from the human data). The bumps in $ \phi_{34} $, accompanying dwells in $ \phi_{13} $, are reproduced when $ a_3\gg a_4 $ (Figure~\ref{fig:triad}B) but not when $ a_3=a_4$ (Figure~\ref{fig:triad}C; see \nameref{section:triad_SI} in \nameref{SI} for more analyses). This example shows that to understand interesting dynamic patterns in specific trials, non-uniform coupling strength is important. 

\section{Discussion}

The present model is a natural generalization of the extended HKB (for $ N=2 $) \cite{Kelso1990dw} to higher dimensions (arbitrary N) and an extension of the classical Kuramoto model (for large $ N $) \cite{Kuramoto1984} to include second-order coupling, thereby reconciling small-scale and large-scale theories of coordination. The model successfully captures key features of multiagent coordination in mid-scale ensembles at multiple levels of description \cite{Zhang2018}. Similar to the HKB model \cite{Haken1985}, second-order coupling is demanded by the experimental observation of antiphase (and associated multistability) but now in eight-person coordination; and similar to the extended HKB \cite{Kelso1990dw}, the model captures how increasing frequency difference $ \delta f $ weakens inphase and antiphase patterns, leading to segregation but now between two groups instead of two persons. This cross-scale consistency of experimental observations may be explained by the scale-invariant nature of the critical coupling ratio $ \kappa_c=1 $, the transition point between monostability (only an all-inphase state) and multistability (states containing any number of antiphase relations). The scale invariance suggests that experimental methods and conclusions for small-scale coordination dynamics have implications for multistability, phase transitions, and metastability at larger scales, and enables a unified approach to biological coordination that meshes statistical mechanics and nonlinear dynamics.

Another generalization of the classical Kuramoto model by Hong and Strogatz \cite{Hong2011} also allows for antiphase-containing patterns (``$ \pi $-state") by letting the sign of the first order coupling ($ a $) be positive for some oscillators (``the conformists") and negative for others (``the contrarians"). However, in contrast to our model, antiphase induced this way does not come with multistability, nor the associated order-to-order transitions observed in human rhythmic coordination \cite{Kelso1995,Fuchs1994}. The second-order coupling in our model allows each individual to be \textit{both} a conformist \textit{and} a contrarian but possibly to different degrees \cite{Kelso2006}. The simple addition of a second stable state may not seem like a big plus at $ N=2 $ (2 stable states), but it rapidly expands the system's behavioral repertoire as the system becomes larger ($ 2^{N-1} $ stable states for $ N $ oscillators; with only first-order coupling, the system always has $ 1^{N-1}=1 $ stable state, and therefore does not benefit from scaling up). This benefit of scale may be how micro-level multistability contributes to the functional complexity of biological systems \cite{Kelso2012,Laurent1999}. 

Outside of the mathematical context of stability analysis, we have to recall that spontaneous social coordination is highly metastable (e.g. Figure~\ref{fig:3df_example_phi}A) \cite{Zhang2018}, captured by the model when frequency diversity is combined with weak coupling (e.g. Figure~\ref{fig:parspace_NHKB}A, in contrast to BC under stronger coupling). Individuals did not become phase-locked in the long term, but coordinated temporarily when passing by a preferred state (inphase and antiphase) \cite{Tognoli2014,Kelso2012} (e.g. red trajectory in Figure~\ref{fig:triad}A). For $ N>2$, an ensemble can visit different spatial organizations sequentially (see examples in \cite{Zhang2018}), forming patterns that extend in both space and time (Figure~\ref{fig:modularfreq} for intragroup patterns), which further expands the repertoire of coordinative behavior (see \nameref{section:meta} in \nameref{SI}). By allowing complex patterns to be elaborated over time, metastability makes a viable mechanism for encoding complex information as real-world complex living systems do (e.g. brain)  \cite{ Kelso1995, Tognoli2014,Breakspear2010,Friston1997,Tognoli2014Frontiers,Ashwin2007,Rabinovich2008}. In contrast, highly coherent patterns like collective synchronization can be less functional and even pathological \cite{Jirsa2014,Tang2017}. Our results call for more attention to these not-quite coherent but empirically relevant patterns of coordination. 

Besides the multistability or multi-clustering in micro patterns (a general feature endowed by higher-order coupling, e.g. \cite{Okuda1993,Hansel1993PA,Hansel1993PRE}), existing mathematical studies suggest that the presence of second-order coupling should also manifest at the macro level in large scale coordination. Naturally, it induces multistability of the order parameter in the thermodynamic limit \cite{Komarov2013,Komarov2014,Vlasov2015,Clusella2016}. It also alters the critical scaling of macroscopic order (see \cite{Strogatz2000} for a summary), i.e. for coupling strength $ K>K_c $ near $ K_c $, the order parameter $ \|H\|$ (norm of the order function \cite{Daido1992}) is proportional to $ (K-K_c)^\beta $, with $ \beta=1/2 $ for the classical Kuramoto model and $ \beta=1 $ when second-order coupling is added \cite{Daido1994, Crawford1999}. For complex biological systems like the brain which appears to operate near criticality \cite{Chialvo2010}, these two types of scaling behavior may have very different functional implications. When modeling empirical data of biological coordination, one may want to have a closer examination or re-examination of the data for multistability and critical scaling of the order parameter, especially if finer level details are not available.

Key experimental observations are captured by our model under the assumptions of uniform coupling (everyone coordinates with others in the same way) and constant natural frequency, but these assumptions may be loosened to reflect detailed dynamics. For example, introducing individual differences in coupling style (equation (\ref{eqn:NHKB_asymcouple})) gives more room to explain how one metastable phase relation may exert strong influence on another (Figure~\ref{fig:triad}A). Long time-scale dynamics observed in the experiment (see \nameref{section:triad_SI} in \nameref{SI}) may also be explained by frequency adaptation, which has been observed in dyadic social coordination \cite{Nordham2018}. A systematic study of the consequences of asymmetric coupling and frequency adaptation on coordination among multiple agents seems worthy of further experimental and theoretical exploration.

To conclude, we proposed a model that captured key features of human social coordination in mid-sized ensembles \cite{Zhang2018}, and at the same time connects well-studied large-scale and small-scale models of biological coordination. The model provides mechanistic explanations of the statistics and dynamics already observed, as well as a road map for future empirical exploration. As an experimental-theoretical platform for understanding biological coordination, the value of the middle scale should not be underestimated, nor the importance of examining coordination phenomena at multiple levels of description.

\section{Materials and Methods}\label{section:methods}
\subsection{Methods of the human experiment}\label{section:humanmethod}
A complete description of the methods of the ``Human Firefly" experiment can be found in \cite{Zhang2018}. Here we only recapitulate a few points necessary for understanding the present paper. For an ensemble of eight people (120 subjects in total), each subject was equipped with a touchpad that recorded his/her tapping behavior as a series of zeros and ones at 250 Hz (1=touch, 0=detach), and an array of eight LEDs arranged in a ring, each of which flashed when a particular subject tapped. For each trial, subjects were first paced with metronomes for 10s, later interacting with each other for 50s (instructed to maintain metronome frequency while looking at others' taps as flashes of the LEDs). Between the pacing and interaction period, there was a 3s transient, during which subjects tapped by themselves. Tapping frequency during this transient has been used to estimate the ``natural frequencies" of the subjects (see \nameref{section:kernel}). During pacing, four subjects received the same metronome (same frequency, random initial phase), and the other four another metronome. The metronome assignments created two frequency groups (say, group $ A $ and $ B $) with intergroup difference $ \delta f=|f_A-f_B|=0 $, $ 0.3 $, or $ 0.6 $ Hz (same average $ (f_A+f_B)/2=1.5 $ Hz). From a single subject's perspective, the LED array looks like the legend of Figure~\ref{fig:3df_example_phi}A (all LEDs emit white light; color-coding only for labeling locations): a subject always saw his/her own taps as the flashes of LED 1, members of his/her own frequency group LED 2-4, and members of the other group LED 5-8 (members from two groups were interleaved to preserve spatial symmetry).

From the tapping data (rectangular waves of zeros and ones), we obtained the onset of each tap, from which we calculated instantaneous frequency and phase. Instantaneous frequency is the reciprocal of the interval between two consecutive taps. Phase ($ \varphi $) is calculated by assigning the onset of the $ n $th tap phase $ 2\pi(n-1) $, then interpolating the phase between onsets with a cubic spline.

\subsection{Estimating the distribution of natural frequencies}\label{section:kernel}

Human subjects have variable capability to match the metronome frequency and maintain it, which in turn affects how they coordinate. To reflect this kind of variability in the simulations, the oscillators' natural frequencies were drawn from a probability distribution around the ``metronome frequency" (central frequencies $ f_A $ and $ f_B $ for groups $ A $ and $ B $). To estimate this distribution from human data, we first approximated the ``natural frequency" of each subject in each trial with the average tapping frequency during the transient between pacing and interaction periods (see \nameref{section:humanmethod}), and subtracted from it the metronome frequency (see blue histogram in Figure~\ref{fig:tranfreq_kernel} from the ``Human Firefly" experiment \cite{Zhang2018}). We then estimated the distribution non-parametrically, with a \textit{kernel density estimator} in the form of 

\begin{align}
\hat{P}(x) = \dfrac{1}{nh}\sum_{i=1}^n K\bigg(\dfrac{x-x
	_i}{h}\bigg)
\end{align}
where the \textit{Kernel Smoothing Function} is Normal, $K(y)=\dfrac{1}{\sqrt{2\pi}}e^{-\frac{y^2}{2}}$. Here $n=2072$ (259 trials $\times$ 8 subjects) from the experiment. We choose the bandwidth $h=0.0219$, which is optimal for a normal density function according to \cite{W.Bowman1997},

\begin{align}
h = \bigg(\dfrac{4}{3n}\bigg)^{1/5}\sigma\label{eqn:kernelh}
\end{align}
where $ \sigma $ is the measure of dispersion, estimated by

\begin{align}
\tilde{\sigma} = \mathrm{median} \{|y_i-\mathrm{median} \{y_i\}|\}/0.6745
\end{align}
where $ y_i $'s are samples \cite{Hogg1979}. The result of the estimation is shown in Figure~\ref{fig:tranfreq_kernel} (red curve).

\subsection{Phase-locking value and level of integration}\label{section:plvdef}
The (short-windowed) phase-locking value (PLV) between two oscillators (say $x$ and $y$) during a trial is defined as

\begin{align}
PLV_{xy}=\dfrac{1}{W}\sum_{w=1}^W \dfrac{1}{M}\bigg| \sum_{m=1}^M \exp(i\phi_{xy}[(w-1)M+m]) \bigg| \label{eqn:plv}
\end{align}
where $\phi_{xy}=\varphi_x-\varphi_y$, $W$ is the number of windows which each $\phi$ trajectory is split into, and $M$ is number of samples in each window (in the present study, $ W=16 $ and $ M=750 $, same as \cite{Zhang2018}).

Intragroup PLV ($PLV_{intra}$) is defined as

\begin{equation}
PLV_{intra}=\bigg(\binom{|A|}{2}+\binom{|B|}{2}\bigg)^{-1} \bigg(\sum_{x,y\in A} PLV_{xy} + \sum_{x,y\in B} PLV_{xy}\bigg)\label{eqn:plvin}
\end{equation}
where $A$ and $B$ are two frequency groups of four oscillators, corresponding to the design of the ``Human Firefly" experiment \cite{Zhang2018}, $A=\{1,2,3,4\}$,  $B=\{5,6,7,8\}$, and $|A|=|B|=4$.

Intergroup PLV ($PLV_{inter}$) is defined as

\begin{equation}
PLV_{inter}=\dfrac{1}{|A||B|} \sum_{x\in A, y\in B} PLV_{xy}. \label{eqn:plvout}
\end{equation}
In both the human and simulated data, comparisons of $ PLV_{intra} $ and $ PLV_{inter} $ for different levels of $ \delta f $ were done using two-way ANOVA with Type III Sums of Squares, and Tukey Honest Significant Difference tests for post-hoc comparisons (shown in Figure~\ref{fig:criticaldf_anova}BD).

The level of integration between two frequency groups is defined based on the relationship between intragroup coordination (measured by $ PLV_{intra} $) and intergroup coordination (measured by $ PLV_{inter} $). The groups are said to be \textit{integrated} if intragroup coordination is positively related to intergroup coordination, and \textit{segregated} if negatively related. Quantitatively, for each combination of intergroup difference $ \delta f $ and coupling strength $ a $ (assuming $ a=b $ for our model, assuming $ b=0 $ for the classical Kuramoto model), we use linear regression 

\begin{align}
PLV_{inter, k}^{(\delta f,a)}=\beta_0^{(\delta f,a)}+\beta_1^{(\delta f,a)} PLV_{intra,k}^{(\delta f,a)}+error_k^{(\delta f,a)}
\end{align} 
where $PLV_{\cdot,k}^{(\delta f, a)}$ is the inter/intra-group PLV for the $k$th trial simulated with the parameter pair $(\delta f, a)$, and the slope of the regression line $\beta_1^{(\delta f,a)}$ is defined as the measure of the \textit{level of integration} between two frequency groups. If $\beta_1>0$, the groups may be said to be integrated; if $\beta_1<0$, segregated. The set $\{(\delta f,a)|\beta_1^{(\delta f,a)}=0\}$ is the \textit{critical boundary} between the domains of intergroup integration and segregation. Theoretical analyses (Section \nameref{section:parspace} in \nameref{SI}) show that this measure is meaningful (i.e. reflecting qualitative differences between dynamics; Figure~\ref{fig:parspace_NHKB}A-C).

\subsection{Method of simulation}
All simulations were done using the Runge-Kutta 4th-order integration scheme, with a fixed time step $ \Delta t=0.004 $ for duration $ T=50 $ (matching the sampling interval and the duration of interaction period of the human experiment \cite{Zhang2018}; second may be used as unit), i.e. for system $ \dot{X}=f(X) $, with initial condition $ X(0)=X_0 $, the $ (n+1) $th sample of the numeric solution can be solved recursively 

\begin{align}
X[n+1]&=X[n]+\dfrac{1}{6}(k_1+2k_2+2k_3+k_4)\\
\intertext{where}
k_1&=\Delta t\, f(X[n])\\
k_2&=\Delta t\, f(X[n]+k_1/2)\\
k_3&=\Delta t\, f(X[n]+k_2/2)\\
k_4&=\Delta t\, f(X[n]+k_3).
\end{align}
The solver was implemented in CUDA C++, ran on a NVIDIA graphics processing unit, solving every 200 trials in parallel for each parameter pair $ (\delta f,a) $. For each trial, initial phases (of eight oscillators) were drawn randomly from a uniform distribution between $ 0 $ and $ 2\pi $, and natural frequencies from distributions defined by equation (\ref{eqn:kernelh}) (reflecting the design of and variability observed in the human experiment \cite{Zhang2018}). Here 200 trials are used per condition, greater than that of the human experiment (see \cite{Zhang2018} and Section \nameref{section:expdesign} in \nameref{SI} for details) to obtain a more accurate estimate of the mean.

\section{Acknowledgments}
The authors of this work are supported by NIMH Grant MH080838, NIBIB grant EB025819, the FAU Foundation, FAU I-SENSE and FAU Brain Institute.

\section{Competing interests}
There are no competing interests.

\bibliography{FireflyModel}


\section{Supplementary Materials}\label{SI}
\beginsupplement
\subsection{Additional analyses on the coexistence of inphase and antiphase preference}\label{section:stats_antiphase}

\begin{figure}[H]
	\begin{fullwidth}
		\centering
		\includegraphics[width=\linewidth]{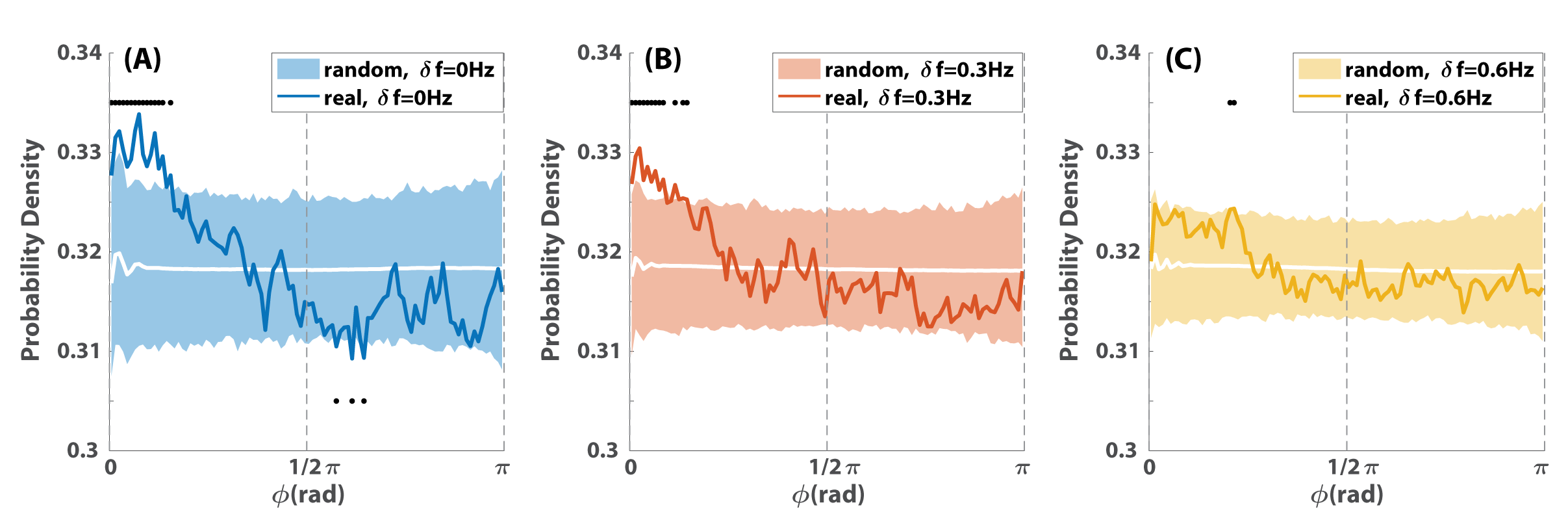}
		\caption{
			Relative phase distributions in the human experiment and comparisons with corresponding chance-level distributions for $ \delta f=0 $ (A), 0.3 (B), and 0.6 Hz (C). Colored solid lines are the probability density functions of all dyadic relative phases for different $ \delta f $'s, each estimated in 100 bins. White solid lines and color-shaded areas are the chance level distributions and corresponding confidence intervals with significance level $ p=0.0005 $ per bin (after Bonferroni correction for $ \hat{p}=0.05 $ for an entire distribution; see the construction of random distributions in Section D in S1 File of \cite{Zhang2018}). Black dots above the distributions mark where the probability density functions are significantly greater than chance, and black dots below mark where they are significantly less than chance (dots appear as bars when significant difference from chance is found in consecutive bins). This is a reproduction of Fig E (B1-B3) in S1 File of \cite{Zhang2018} but with all bins significantly different from chance marked (rather than as in \cite{Zhang2018}, bins were marked only if significance was found in 3 or more consecutive bins).
		}\label{fig:phiDist_wCI}
	\end{fullwidth}
\end{figure}

The coexistence of inphase and antiphase preference in human coordination \cite{Zhang2018} (Figure~\ref{fig:3df_example_phi}D-F) and model behavior (Figure~\ref{fig:modeldyn_phi}D-F) is reflected by the location of troughs (minima in the probability density functions) separating the inphase and antiphase peaks. In the human data (Figure~\ref{fig:3df_example_phi}D-F), the minima are at $ \phi= 0.62\pi$ , $ 0.77\pi $, and $ 0.8\pi $ (away from both $ 0 $ and $ \pi $) for $ \delta f= 0$, $ 0.3 $, and $ 0.6 $ Hz respectively, with the minimum for $ \delta f=0 $ Hz significantly less than chance ($ p<0.0005 $, Figure~\ref{fig:phiDist_wCI}A; nowhere with probability density significantly less than chance for $ \delta f= 0.3$, and $ 0.6 $ Hz, as shown in Figure~\ref{fig:phiDist_wCI}BC). This suggests that there is an unstable phase relation between inphase and antiphase, which is most prominent for $ \delta f=0 $ Hz. This is well reflected in the behavior of the present model (i.e. equation (\ref{eqn:uniformNHKB}) with $ a=b=0.105 $) shown in Figure~\ref{fig:modeldyn_phi}D-F, where the minima of the probability density functions are at $ \phi=0.6\pi $, $ 0.7\pi $, and $ 0.67\pi $ for $ \delta f=0 $, $ 0.3 $, and $ 0.6 $ Hz and the contrast between the minimum and the antiphase peak is most prominent for $ \delta f=0 $ Hz (Figure~\ref{fig:modeldyn_phi}D). On the other hand, for the Kuramoto model (i.e. equation (\ref{eqn:uniformNHKB}) with $ b=0 $), the minima of the probability density functions are always at $ \phi=\pi $ for all $ \delta f $'s (Figure~\ref{fig:modeldyn_phi}G-I), reflecting the instability related to antiphase when second order coupling is removed (supported by analytical results in Section \nameref{section:CB_multistability}). Thus, the experimental phenomena from \cite{Zhang2018} cannot be fully captured without the second order coupling.

\subsection{Choosing the appropriate coupling strength}\label{section:parspace}
What we want to see is how the present model behaves as we manipulate the diversity of natural frequency $ \omega_i $'s just as we did to human subjects. However, there remain two unknown parameters to be taken care of, namely the coupling stength $ a $ and $ b $ in equation (\ref{eqn:uniformNHKB}). Before systematically finding the appropriate coupling strength, we want to first show qualitatively how it affects the dynamics. 

\begin{figure}[H]
	\begin{fullwidth}
		\centering
		\includegraphics[width=\linewidth]{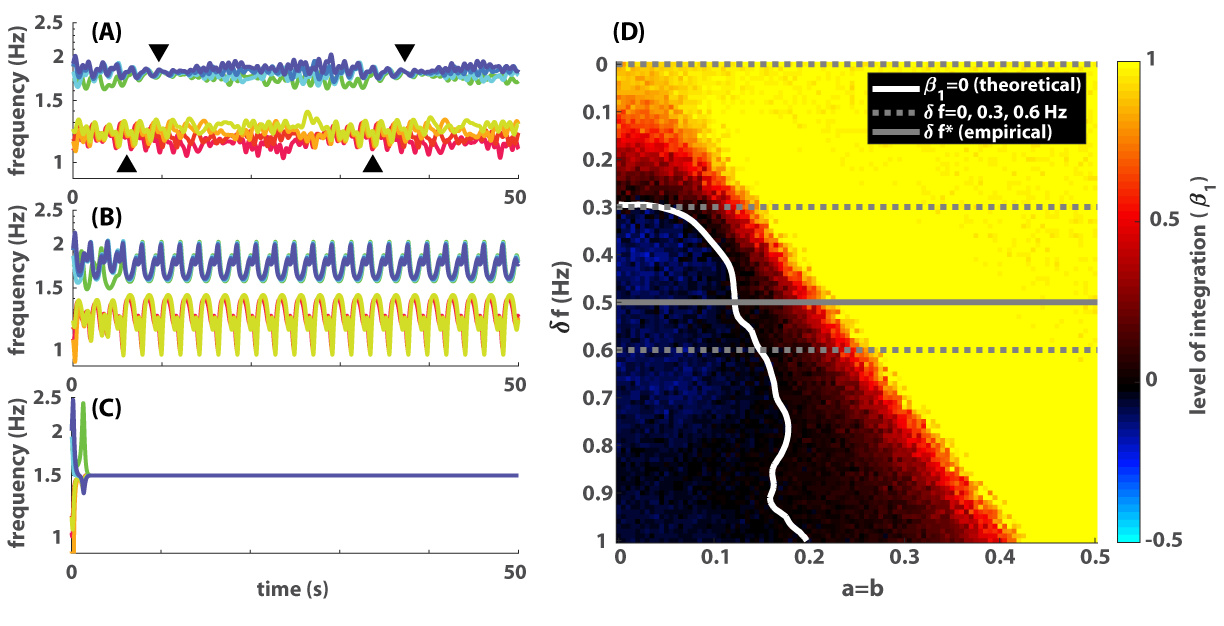}
		\caption{The effect of diversity and coupling strength on the level of integration between groups. (A-C) show frequency dynamics of three simulated trials, with increasing coupling strength ($ a=b=0.1,\,0.2,\,0.4 $ respectively) and all other parameters identical (the warm-color group's natural frequencies evenly spread in the interval $ [f_A-0.08 Hz,f_A+0.08 Hz] $ with $ f_A=1.2 $ Hz, similarly for the cold-color group in $ [f_B-0.08 Hz,f_B+0.08 Hz] $ with $ f_B=1.8 $ Hz; initial phases are random across oscillators but the same across trials). When the coupling is too strong (C), all oscillators lock to the same steady frequency. When the coupling is moderate (B), oscillators split into two frequency groups, phase-locked within themselves, interacting metastably with each other (dwell when trajectories are close, escape when trajectories are far apart). When the coupling is weak (A), intragroup coordination also becomes metastable seen as episodes of convergence (black triangles) and divergence. (D) shows the level of intergroup integration quantitatively ($ \beta_1 $, color of each pixel) for each combination of frequency diversity $ \delta f $ and coupling strength $ a=b $. White curve indicates the critical boundary between segregation (blue area on the left, $ \beta_1<0 $) and integration (red and yellow area on the right, $ \beta_1>0 $). Within the regime of integration, the yellow area indicates complete integration ($ \beta_1\approx 1 $) where there is a high level of phase locking, and the red area indicates partial integration ($ 0<\beta_1\ll 1 $) suggesting metastability. Dashed gray lines label $ \delta f $'s that appeared in the human experiment. Solid gray line labels the empirically estimated critical diversity. }\label{fig:parspace_NHKB}
	\end{fullwidth}
\end{figure}

Three simulated trials with increasing coupling strength are shown in Figure~\ref{fig:parspace_NHKB} from A to C, whereas the initial phases and natural frequencies are the same across trials (warm-color group centered around $ f_A=1.2$ Hz, cold-color group $ f_B=1.8$ Hz, corresponding to the condition $ \delta f=0.6 $ Hz). When the coupling is weak ($ a=b=0.1$, Figure~\ref{fig:parspace_NHKB}A), oscillators are well-segregated into two frequency groups. Within each frequency group, members intermittently converge (marked by black triangles) then diverge, reflecting metastability at a group level (collective dwells). For intermediate coupling ($ a=b=0.2 $, Figure~\ref{fig:parspace_NHKB}B), oscillators within each group are locked together, interacting strongly as a whole with the other frequency group (seen as the oscillation of frequency), so that the ensemble ($ N=8 $) behaves like a dyad ($ N=2 $). Finally, for strong coupling ($ a=b=0.4 $, Figure~\ref{fig:parspace_NHKB}C), everyone converges to a single steady frequency. We see a progression from group-level segregation to integration from (A) to (C), indicating the important role of coupling strength in determining intergroup relation. Qualitatively, the model's behavior under weak coupling (Figure~\ref{fig:parspace_NHKB}A) is closer to human behavior (Figure~\ref{fig:3df_example_phi}C) than that of stronger coupling. Next we take a more quantitative look. 

\noindent
To quantify the joint effect of frequency diversity ($ \delta f $) and coupling strength ($ a=b $ for simplicity) on integration and segregation between two frequency groups, we calculated the level of intergroup integration ($ \beta_1 $) for simulated trials using the same method as for the human experiment (see \nameref{section:plvdef} in \nameref{section:methods} in the main text). For each parameter pair $ (\delta f, a) $ with $ a=b $, we simulated 200 trials. In each simulated trial, two frequency groups $ A $ and $ B $ each consists of four oscillators ($ \varphi_1,\cdots ,\varphi_4 $ in group $ A $, $ \varphi_5,\cdots,\varphi_8 $ in group $ B $). The natural frequency of oscillators in group $ A $ (i.e. $ \omega_1,\cdots,\omega_4 $, divided by $ 2\pi $) was drawn from a distribution $ P(f_A) $ centered around $ f_A $ (corresponds to the metronome frequency for the group in the human experiment), and $ P(f_B) $ for group $ B $. The difference between two groups $ \delta f= |f_A-f_B|$ corresponds to the level of diversity in the human experiment. Here the probability density function $ P(f) $, which defines frequency dispersion within each group, was obtained by an nonparametric estimation of the empirical distribution (see \nameref{section:methods} in the main text).

The level of intergroup integration for simulated trials is shown in Figure~\ref{fig:parspace_NHKB}D as the color of each pixel (diversity $ \delta f $ as y-coordinate; coupling strength $ a=b $ as x-coordinate). Three regimes are apparent: the highly integrated (yellow, $ \beta_1 \approx 1$), the partially integrated (red, $ 0<\beta_1\ll 1 $), and the segregated (blue, $ \beta_1<0 $). Between the red and blue area is the critical boundary (white solid line, $ \beta_1=0 $), separating the regimes of integration and segregation. With any fixed coupling strength, for the critical boundary to fall between $ \delta f=0.3 $ Hz and $ \delta f=0.6 $ Hz as in the human experiment, the coupling strength has to be weak (for $ \delta f=0.6 $ Hz, $ \beta_1<0 $ only when $ a=b<0.15 $ ) but not too weak (for $ \delta f=0.3 $ Hz, $ \beta_1>0 $ only when $ a=b>0.05 $). Without risking overfitting, we simply choose the coupling strength $ a=b=0.105 $, for which the level of intergroup integration is the closest to experimental observation for $ \delta f=0.3 $ Hz ($ \beta_1=0.31 $). 

\subsection{Empirical distribution of tapping frequency around metronome frequency}\label{section:empiricaldistribution}
\begin{figure}[H]
	\centering
	\includegraphics[width=1\textwidth]{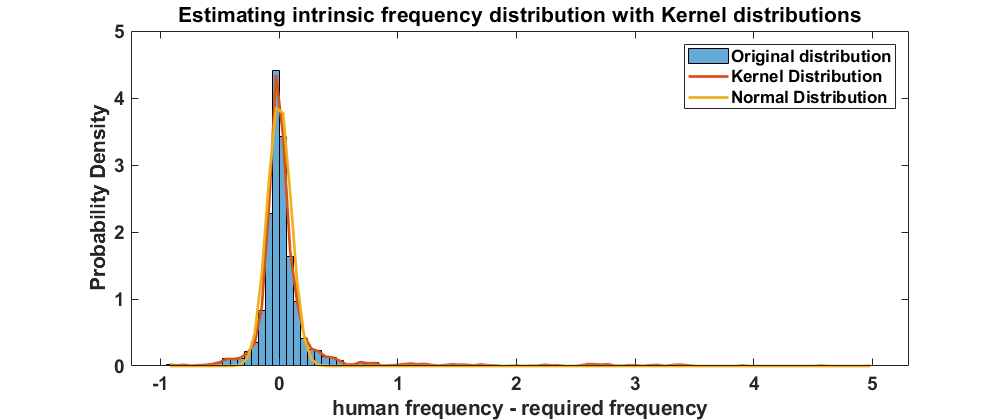}
	\caption{Distribution of human movement frequency around metronome frequencies and its estimation.}
	\label{fig:tranfreq_kernel}
\end{figure}

In the ``Human Firefly" experiment \cite{Zhang2018}, subjects' tapping frequency during the transient between pacing and interaction (a proxy to ``natural frequency"; see \nameref{section:methods} in main text) dispersed around the metronome frequencies. The distribution of this deviation from metronome frequencies is shown in Figure~\ref{fig:tranfreq_kernel} (blue histogram). Most of the time, subjects were very close to the metronome frequency (peak around zero). We can use a normal distribution $ \N(\mu,\sigma) $ to capture this peak (Figure~\ref{fig:tranfreq_kernel} yellow line), where parameters $ \mu=0 $ and $ \sigma=0.0986 $ (Hz) were estimated using the median and $ 10 $th percentile of the empirical distribution. We can see a difference between the empirical distribution and the normal distribution - the normal distribution (yellow line) does not capture the fat-tails of the empirical distribution (blue bars exceed yellow line on its shoulders). These ``mutant fireflies" making up the fat-tails are not to be dismissed as out-liers, because they contribute to the behavior of others in the ensemble. To better represent the empirical distribution, we used Kernel Density Estimation (with a normal kernel) as described in section \nameref{section:kernel} of \nameref{section:methods} in the main text, and the result of estimation is shown as the red line in Figure~\ref{fig:tranfreq_kernel} (named kernel distribution). The kernel distribution better captures the tails of the empirical distribution and was used to generate natural frequencies of oscillators in the simulations. 

\subsection{Examples of dynamics with intergroup coupling removed}\label{section:modulardynamics}
\begin{figure}[H]
	\centering
	\includegraphics[width=.8\textwidth]{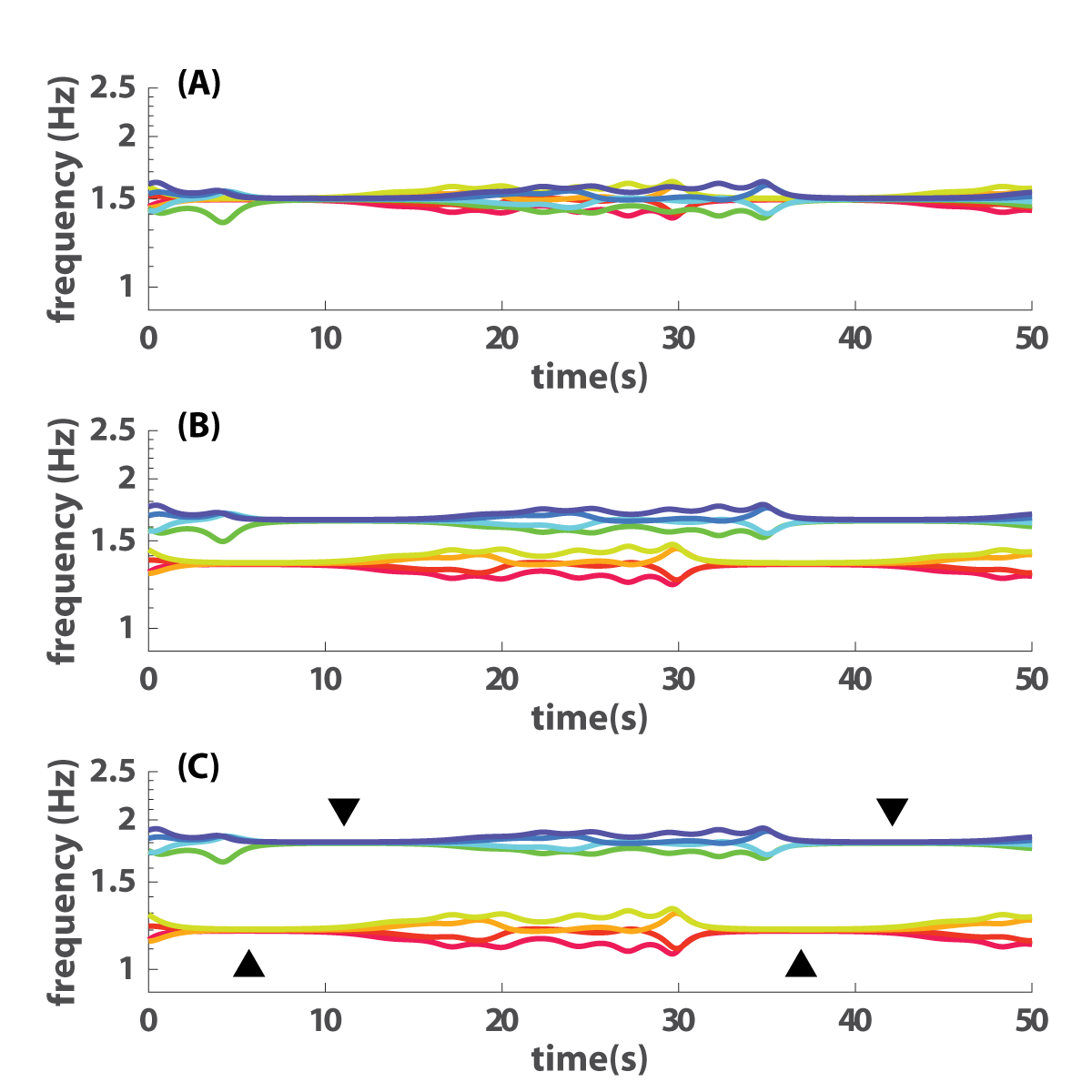}
	\caption{Intragroup dynamics without intergroup coupling, for intergroup difference $ \delta f=0 $ (A), $ \delta f=0.3 $ (B), and $ \delta f=0.6 $ Hz (C). }\label{fig:modularfreq}
\end{figure}

By removing intergroup coupling, we obtain a modification of equation (\ref{eqn:uniformNHKB})

\begin{align}
\dot{\varphi}_{i}=\omega_i-a\sum_{j=1}^{N}e_{ij}\sin \phi_{ij}-b\sum_{j=1}^{N}e_{ij}\sin 2\phi_{ij}\label{eqn:rmintergroupcouple}
\end{align}
where $ e_{ij}=1 $ if $ i,j\in \{1,2,3,4\} $ or $ i,j\in\{5,6,7,8\} $, $ e_{ij}=0 $ otherwise, for $ N=8 $. The resulting dynamics (with all other parameters the same as examples in Figure~\ref{fig:modeldyn_phi}A-C in the main text) are given in Figure~\ref{fig:modularfreq}. Within each frequency group (one group in cold colors, one group in warm colors), we see the same intragroup metastable dynamics being repeated regardless of intergroup difference ($ df=0,0.3,0.6 $ Hz for Figure~\ref{fig:modularfreq}A, B, C respectively). These trials, without intergroup coupling, provide a baseline dynamics for comparison with Figure~\ref{fig:modeldyn_phi}A-C, revealing the effect of intergroup influence. It turns out that for a given intragroup coupling, intragroup metastability comes from intragroup dispersion of natural frequencies. Metastability vanishes when two metastable groups have no intergroup difference (Figure~\ref{fig:modeldyn_phi}A). In other words, without intergroup difference ($ \delta f=0 $), there are more oscillators within the same range of frequency, which cooperatively increases intragroup coordination. If we remove this intragroup dispersion of natural frequency (along with the metastability), we can no longer reproduce the experimental observation that intragroup coordination was weakened and altered by intergroup differences (see \nameref{section:noingroupvarfreq}  for a statistical analysis).

\begin{figure}[H]
	\centering
	\includegraphics[width=0.8\textwidth]{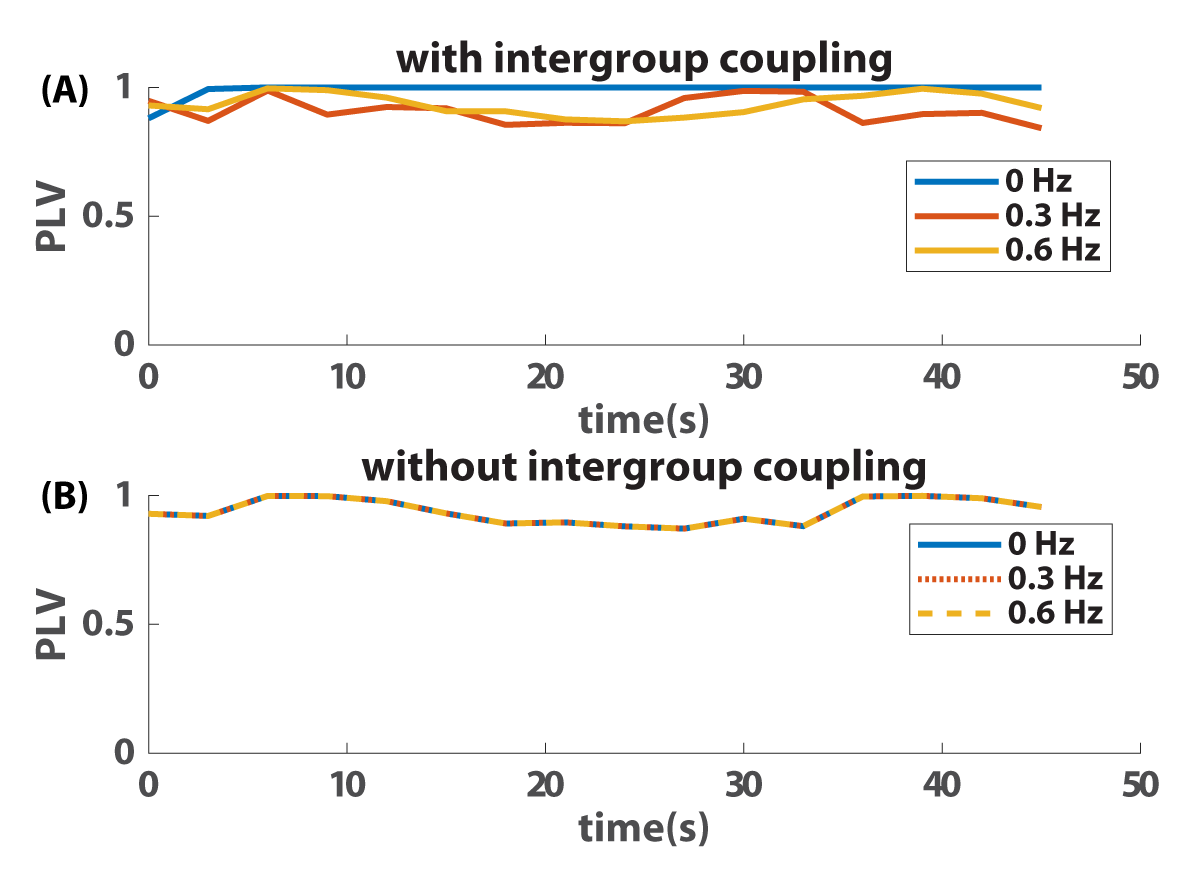}
	\caption{Dynamics of intragroup phase-locking value (PLV) with (A) and without (B) intergroup coupling.}\label{fig:modularplv}
\end{figure}

With intergroup coupling, the time scale of metastability is modified by $ \delta f $, as shown in Figure~\ref{fig:modeldyn_phi}BC where the interval between two episodes of convergence (black triangles) is shorter for $ \delta f=0.3 $ Hz (B) than for $ \delta f=0.6 $ Hz (C). In Figure~\ref{fig:modularplv}A, this is also visualized as the dynamics of phase-locking value (PLV) within groups (average PLV of all intragroup dyads in 3-s windows). When oscillators within groups converge, PLV is close to 1, and the interval between two consecutive peaks in a PLV trajectory reflects the time scale of the metastable coordination. Without influence from the other group, the time scales are exactly the same (trajectories exactly on top of each other in Figure~\ref{fig:modularplv}B). With influence from the other group, the time scale depends on the level of intergroup difference (inter-peak intervals for $ \delta f=0.3 $ Hz was much shorter than that of $ \delta f=0.6 $ Hz in Figure~\ref{fig:modularplv}A). Perhaps, we can consider $ \delta f=0 $ Hz (i.e. lost of metastability, \ref{fig:modularplv}A blue line) as the special case where the inter-convergence interval is zero.  

It is also interesting to notice that for $ \delta f=0.6 $ Hz, the metastable time-scale of the trial with intergroup coupling (Figure~\ref{fig:modularplv}A yellow line) is very similar to that of the trial without intergroup coupling  (Figure~\ref{fig:modularplv}B yellow line). This may be connected to the fact that $ \delta f=0.6 $ Hz (given $ a=b=0.105 $) is in the regime of intergroup segregation. It is perhaps a hypothesis worth further investigation that the level of intergroup integration (as measured by $ \beta_1 $, see main text) reflects how the time scale of intragroup metastability was affected by intergroup difference.  Here our discussion on these examples is only to provide an intuitive understanding of the dynamics. 

\subsection{Effect of reduced intragroup variability in natural frequency}\label{section:noingroupvarfreq}
\begin{figure}[H]
	\begin{fullwidth}
		\centering
		\includegraphics[width=\linewidth]{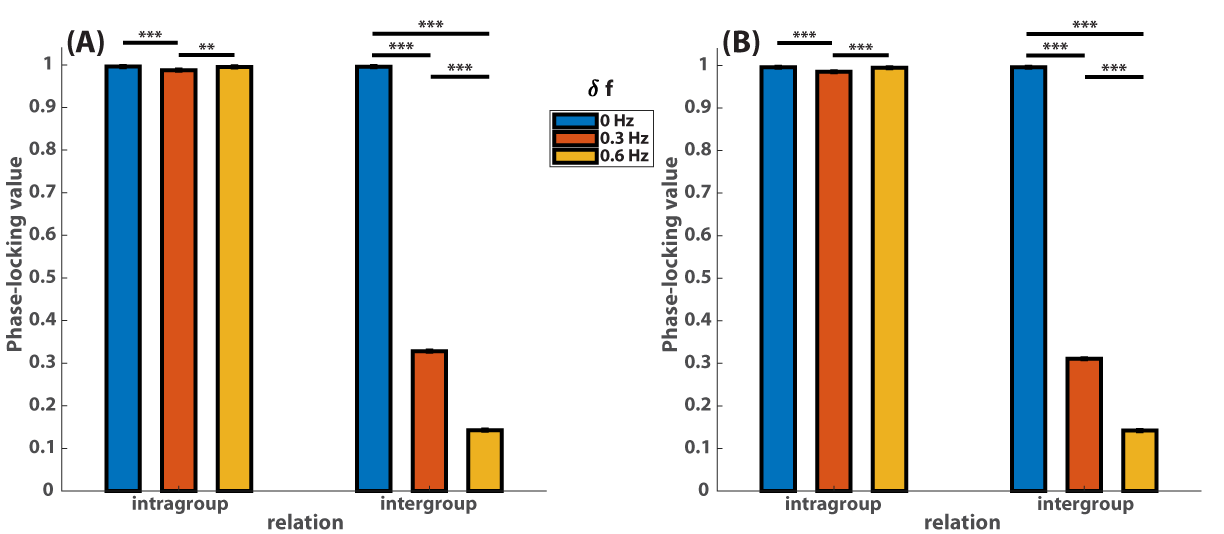}
		\caption{Intragroup and intergroup phase-locking by different levels of diversity $ \delta f $ for simulated data with identical natural frequency within groups. (A) shows the results of simulations with uniform coupling, and (B) non-uniform coupling ($ a $'s and $ b $'s are randomly distributed in the interval $ [0,0.2] $ see text for details).  }\label{fig:symbreak}
	\end{fullwidth}
\end{figure}

Recall that the reduction in intragroup coordination shown in Figure~\ref{fig:criticaldf_anova}D (left) was based on simulations with nontrivial dispersion in natural frequency within each group, reflecting the natural variability carried into the experiment by human subjects. What if we remove that intragroup dispersion? As shown in Figure~\ref{fig:symbreak}A (left three bars), intragroup coordination becomes all very close to the maximal level (phase-locking value close to 1) for all diversity conditions (MANOVA interaction effect $ F(2,19194)=50152 $, $ p<0.001 $); we no longer see the large drop in intragroup coordination as seen in Figure~\ref{fig:criticaldf_anova}BD. Even if we break the symmetry in coupling strength (use equation (\ref{eqn:phaseorder2}) with random coefficients, instead of uniform coupling in equation (\ref{eqn:uniformNHKB}); see \nameref{section:random_a}  for details), the phenomenon is not recovered (Figure~\ref{fig:symbreak}B very similar to A; MANOVA interaction effect $ F(2,19194)=59678 $, $ p<0.001 $). By studying the model's behavior, we found that the reduction in intragroup coordination due to intergroup difference, as observed in the human experiment, mainly depends on asymmetry in natural frequency rather than coupling strength. 

\subsection{Random coupling}\label{section:random_a}
To study the effect of symmetry breaking in coupling strength, we generated random coefficients for equation (\ref{eqn:phaseorder2}), following a uniform distribution on the interval $ [0,a_{max}] $,

\begin{align}
P(a)= \dfrac{1}{a_{max}}.
\end{align}
We simulated 200 trials for each parameter pair $ (\delta f=0.3Hz, a_{max}) $ for $ a_{max}\in [0,1] $ (discretized into intervals of length 0.01) with initial phases randomly distributed from $ 0 $ to $ 2\pi $ and natural frequencies following the empirical distribution from the human experiment (see \nameref{section:empiricaldistribution}). We then find the value of $ a_{max} =0.2$, which produces the level of intergroup integration ($ \beta_1 $) closest to the experimental value (0.31). Using this fitted $ a_{max} $, we simulated 200 trials with no intragroup dispersion in natural frequency, which were used to produce results in Figure~\ref{fig:symbreak}B.

\subsection{Intergroup relation without second order coupling}\label{section:parspace_kura}
\begin{figure}[H]
	\centering 
	\includegraphics[width=0.7\textwidth]{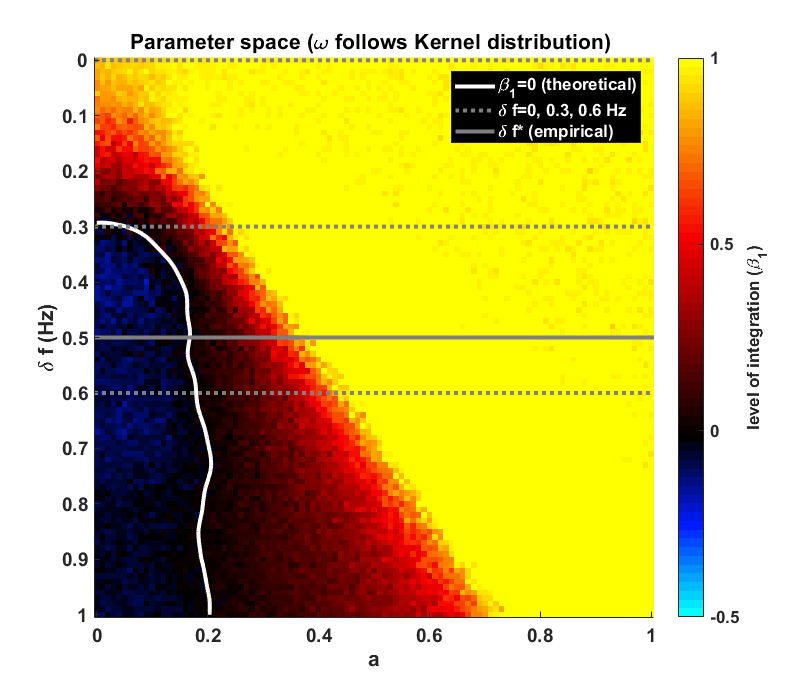}
	\caption{Level of integration between groups by $ \delta f $ and $ a $, with $ b=0 $.} \label{fig:kura_parspace}
\end{figure}

\begin{figure}[H]
	\begin{fullwidth}
		\includegraphics[width=\linewidth]{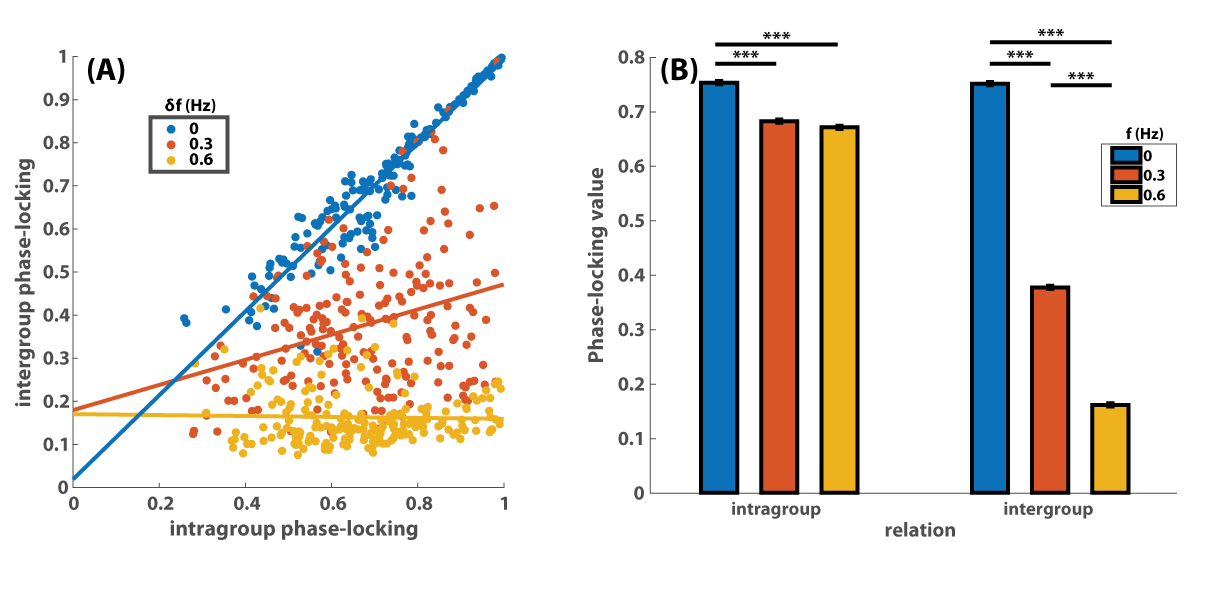}
		\caption{Intragroup, intergroup coordination and the relationship between them for $ a=0.154 $ and $ b=0 $. Here the level of coordination is measured by phase-locking value (see main text for definitions). (A) shows the relationship between intragroup (x-coordinate of each dot) and intergroup coordination (y-coordinate of each dot) for different levels of diversity (color code). The solid lines are corresponding regression lines whose slope quantifies the level of integration between two frequency groups. (B) shows the average intragroup (left three bars) and intergroup coordination (right three bars) for different levels of diversity (color code).}\label{fig:kura_regression}
	\end{fullwidth}
\end{figure}

To examine whether the second order coupling term (i.e. $ b\sum\sin 2\phi_{ij} $) in equation (\ref{eqn:uniformNHKB}) is necessary for reproducing key experimental results, we let $ b=0 $ and followed the exact same analysis as for the case of $ b\neq 0 $. The results are shown in Figure~\ref{fig:kura_parspace} (its $ b\neq 0 $ counterpart is Figure~\ref{fig:parspace_NHKB}D), and Figure~\ref{fig:kura_regression}AB (its $ b\neq 0 $ counterpart is Figure~\ref{fig:criticaldf_anova}CD). \\
Figure~\ref{fig:kura_parspace} shows the organization of the parameter space $ \delta f\times a $ in terms of the level of integration between groups ($ \beta_1 $, see definition in main text). Similar to Figure~\ref{fig:parspace_NHKB}D (for $ b\neq 0 $), the space consists of three regions - complete integration ($ \beta_1\approx 1 $, yellow), partial integration ($ 0<\beta_1\ll 1 $, red), and segregation ($ \beta_1<0 $, blue) - arranged from upper right to lower left. Figure~\ref{fig:kura_parspace} is approximately a scaled version of Figure~\ref{fig:parspace_NHKB}D along $ a $. 

We estimated the coupling strength to be $ a=0.154 $, where the corresponding level of intergroup integration for $ \delta f=0.3 $ Hz is the closest to the empirical value (up to $ 10^{-3} $ precision for a; for $ a=0.154$, $ \beta_1(0.3 Hz)=0.29 $, the empirical value is $ 0.31 $). The corresponding relations between intragroup and intergroup coordination is shown in Figure~\ref{fig:kura_regression}A and average intra/intergroup coordination in Figure~\ref{fig:kura_regression}B for different levels of $ \delta f $. 

In Figure~\ref{fig:kura_regression}A, each dot represents a particular trial with its x-coordinate indicating the average intragroup coordination (measured by phase-locking value, see \nameref{section:methods} in main text) and y-coordinate the average intergroup coordination, whereas the color indicates the diversity $ \delta f $. Similar to the human experiment and the case of $ b\neq 0 $, more intragroup coordination is associated with more intergroup coordination (i.e. intergroup integration) for $ \delta f=0 $ and $ 0.3 $ Hz (blue, red regression lines with positive slopes), and less intergroup coordination (i.e. intergroup segregation) for $ \delta f=0.6 $ Hz (yellow regression line with negative slope). Two differences are (1) the $ \beta_1 $ for $ \delta f=0.6 $ Hz and $ b=0 $ is not significantly different from zero ($ p>0.05 $; see main text for more statistics), where as its counterparts in the human data and the case of $ b\neq 0 $ are ($ p<0.05 $); (2) in the human data and the case of $ b\neq 0 $, three regression lines intersect at almost the same point (see Figure~\ref{fig:criticaldf_anova}A, C), which is not the case for $ b=0 $ (Figure~\ref{fig:kura_regression}A). 

In Figure~\ref{fig:kura_regression}B, we show the average level of intragroup and intergroup coordination (again, in terms of phase-locking values). Intragroup coordination is reduced by the presence of intergroup difference (red, yellow bars on the left significantly shorter than blue bar). Intergroup coordination is more dramatically reduced by intergroup difference. Overall, these results resemble those of the human data and the case of $ b\neq 0 $.

\subsection{Multistability of the present model} \label{section:CB_multistability} 

The equations for $N$ Kuramoto oscillators with the same natural frequency, coupled to one another with a uniform coupling $a > 0$ are 
\begin{equation}\label{CB:KMdef}
\dot\varphi_i = - a \sum_j \sin(\varphi_i - \varphi_j).
\end{equation}
These equations can be recast in the mean-field form 
\begin{equation}\label{CB:KMord}
\dot\varphi_i = - r \sin(\varphi_i - \psi) 
\qquad\text{with}\qquad 
r\, \ee^{\ii\psi} := a \sum_j \ee^{\ii\varphi_j}, 
\end{equation}
which admit two types of fixed points: either \textbf{(a)} $r = 0$, or \textbf{(b)} $\sin (\varphi_i - \psi) = 0$ for each oscillator.  In either case, the linearized equations governing the evolution of a small perturbation $\delta\varphi$ away from a fixed point are 
\begin{align}\label{CB:KMlin}
\delta\dot\varphi_i 
&= - a \sum_j \cos(\varphi_i - \varphi_j)\, (\delta\varphi_i - \delta\varphi_j)
\notag\\
&= - r \cos(\varphi_i - \psi)\, \delta\varphi_i 
+ a \sum_j \cos(\varphi_i - \varphi_j)\, \delta\varphi_j
\end{align}
We will study these linearized equations in the two cases separately.

\paragraph{Case (a)}
The first term of equation (\ref{CB:KMlin}) vanishes in this case.  If we assume further that $\delta\varphi_j = 0$ initially for all but one oscillator, then the simplified equations are 
\begin{equation}\label{CB:KMlina}
\delta\dot\varphi_i = a \cos(\varphi_i - \varphi_j)\, \delta\varphi_j.
\end{equation}
In particular, $\delta\varphi_j$ itself grows exponentially at a rate $a$, so this fixed point cannot be stable.

\paragraph{Case (b)}
In this case there are two subgroups of oscillators, all exactly inphase within their group, and exactly antiphase to the other group.  These groups cannot be equal in number because then $r = 0$ in equation (\ref{CB:KMlin}).  Accordingly, we have $r = (n_+ - n_-) a$, where $n_+ > n_-$ are the sizes of the in- and antiphase groups (relative to the mean oscillator $\psi$, which of course is inphase with the larger group).  The linearized equations become 
\begin{equation}\label{CB:KMlinb}
\delta\dot\varphi_i = - (n_+ - n_-) a s_i\, \delta\varphi_i 
+ a s_i \sum_j s_j\, \delta\varphi_j, 
\end{equation}
where $s_i := \cos(\varphi_i - \psi) = \pm 1$ indicates whether $\varphi_i$ is in- or antiphase to $\psi$.  This equation can be recast in the matrix form 
\begin{equation}\label{CB:KMJac}
\delta\dot\varphi 
= J\, \delta\varphi 
:= \bigl[ a SHS - (a \tr S) S \bigr]\, \delta\varphi, 
\end{equation}
where $S$ is the $N \times N$ matrix with non-zero entries $s_i = \pm 1$ along the diagonal and $H$ is the $N \times N$ matrix with all entries equal to $+1$.  Our goal is to show that the Jacobian matrix $J$ has at least one positive eigenvalue.  This will imply that the fixed point is unstable dynamically.  To do this, first we use the elementary identities $S^2 = 1$ and $H^2 = (\tr 1) H$ to calculate 
\begin{equation}\label{CB:KMJ^2}
J^2 = (a \tr 1) J 
- (a \tr S) \bigl[ a HS + a SH - (a \tr 1) S - (a \tr S) 1 \bigr], 
\end{equation}
where $1$ denotes the $N \times N$ identity matrix.  The additional identity $HSH = (\tr S) H$ then gives 
\begin{equation}\label{CB:KMJ^3}
J^3 = (a \tr 1) J^2 + (a \tr S)^2 J 
+ (a \tr S)^2 \bigl[ a H - (a \tr 1) 1 \bigr]
\end{equation}
Applying all three of these identities one last time yields 
\begin{equation}\label{CB:KMJ^4}
J^4 = (a \tr 1) J^3 + (a \tr S)^2 J^2 - (a \tr 1) (a \tr S)^2 J.
\end{equation}
That is, $J$ solves a quartic polynomial, which moreover factors in the form 
\begin{equation}\label{CB:KMpol}
J \bigl[ J - (a \tr 1) 1 \bigr] \bigl[ J - (a \tr S) 1 \bigr] \bigl[ J + (a \tr S) 1 \bigr] = 0.
\end{equation}
This is clearly the minimal-order polynomial that $J$ solves, and it has all distinct roots.  It follows that $J$ has a complete basis of eigenvectors with eigenvalues $\lambda_0 := 0$, $\lambda_* := N a$, $\lambda_+ := (n_+ - n_-) a$, and $\lambda_- := (n_- - n_+) a$.  (We can't tell the multiplicity of each of these eigenvalues from this calculation, but each has at least a one-dimensional eigenspace associated to it.)  The zero eigenvalue arises because the right side of equation (\ref{CB:KMdef}) involves only \emph{relative} phases, so the dynamics is insensitive to rigid rotations $\varphi_i \mapsto \varphi_i + \theta$ for all $i$.  The eigenvalues $\lambda_*$ and $\lambda_+$, meanwhile, are strictly positive, and show that this fixed point is unstable.

The lone exception to this argument occurs when $n_- = 0$, and therefore $S = 1$.  Then we have $J = a [H - N 1]$, which clearly has a zero eigenspace corresponding to the rigid rotation of all oscillators in the system (\textit{i.e.}, all $\delta\varphi_i$ equal to one another).  Apart from this, there is only a single, complementary eigenspace of dimension $N - 1$ associated with the eigenvalue $\lambda = - N a$.  The configuration with all oscillators exactly inphase is therefore the \emph{only} stable fixed point solution of the Kuramoto model.

Our model (with second-order coupling), on the other hand, has multiple stable fixed points for suitable values of its parameters.  For uniformly coupled, identical oscillators, our equations are 
\begin{equation}\label{CB:HKBdef}
\dot\varphi_i 
= - a \sum_j \sin(\varphi_i - \varphi_j) 
- b \sum_j \sin 2(\varphi_i - \varphi_j).
\end{equation}
The fixed points of the Kuramoto model with each $\varphi_i$ equal either to $\psi$ or to $\psi + \pi$ are also fixed points of these equations, and the linearized equations around such a fixed point are 
\begin{equation}\label{CB:HKBlin}
\delta\dot\varphi_i 
= - (n_+ - n_-) a s_i\, \delta\varphi_i
+ a s_i \sum_j s_j\, \delta\varphi_j
- 2 N b\, \delta\varphi_i 
+ 2 b \sum_j \delta\varphi_j.
\end{equation}
Here again we set $s_i := \cos(\varphi_i - \psi) = \pm 1$ and let $n_\pm$ denote the numbers of oscillators with $s_i = \pm 1$.  The matrix form of these linearized equations is 
\begin{equation}\label{CB:HKBJac}
\delta\dot\varphi 
= J\, \delta\varphi 
:= \bigl[ a SHS - (a \tr S) S + 2b H - (2b \tr 1) 1 \bigr]\, \delta\varphi, 
\end{equation}
where the matrices $S$, $H$, and $1$ are defined as before.  This Jacobian matrix differs from the Kuramoto Jacobian, which now we denote $J_a$, by its last two terms.  Importantly, we have 
\begin{equation}\label{CB:HKBJaH}
J_a H = a SHSH - (a \tr S) H = 0 
\end{equation}
because $HSH = (\tr S) H$.  It follows that all cross-terms vanish in any binomial expansion:  
\begin{equation}\label{CB:HKBJ^n}
\bigl[ J + (2b \tr 1) 1 \bigr]^n 
= \bigl[ J_a + 2b H \bigr]^n 
= J_a^n + \bigl[2b H \bigr]^n 
\end{equation}
for all integers $n$.  Applying these results in equation (\ref{CB:KMJ^3}) then shows that 
\begin{multline}\label{CB:HKBcub}
\bigl[ J + (2b \tr 1) \bigr]^3 
- (a \tr 1) \bigl[ J + (2b \tr 1) \bigr]^2 
- (a \tr S)^2 \bigl[ J + (2b \tr 1) \bigr] 
+ (a \tr S)^2 (a \tr 1) 1 
\\
= (a \tr S)^2 a H 
+ \bigl[ 2b H \bigr]^3 
- (a \tr 1) \bigl[ 2b H \bigr]^2 
- (a \tr S)^2 \bigl[ 2b H \bigr].
\end{multline}
Each term on the right here vanishes if we multiply through by $J$.  Meanwhile, the cubic polynomial on the left is the same one from the Kuramoto case, with its argument shifted by $J \mapsto J + (2b \tr 1) 1$.  It factors in the same way as before to give the minimal polynomial 
\begin{equation}\label{CB:HKBpol}
J 	\bigl[ J + (2b \tr 1 - a \tr 1) 1 \bigr] 
\bigl[ J + (2b \tr 1 - a \tr S) 1 \bigr] 
\bigl[ J + (2b \tr 1 + a \tr S) 1 \bigr] = 0
\end{equation}
for the present model.  The \emph{non-zero} eigenvalues of the Kuramoto models are therefore all shifted by the same amount, giving $\lambda_0 = 0$, $\lambda_* = (a - 2b) N$, and $\lambda_\pm := \pm (n_+ - n_-) a \tr S - 2Nb$.  These are all negative as long as $2b > a$, the same condition that governs the HKB model for dyadic coordination.  Our model is multistable when its parameters satisfy this condition.
\subsection{Additional triadic dynamics}\label{section:triad_SI}
\begin{figure}[H]
	\begin{fullwidth}
		\includegraphics[width=\linewidth]{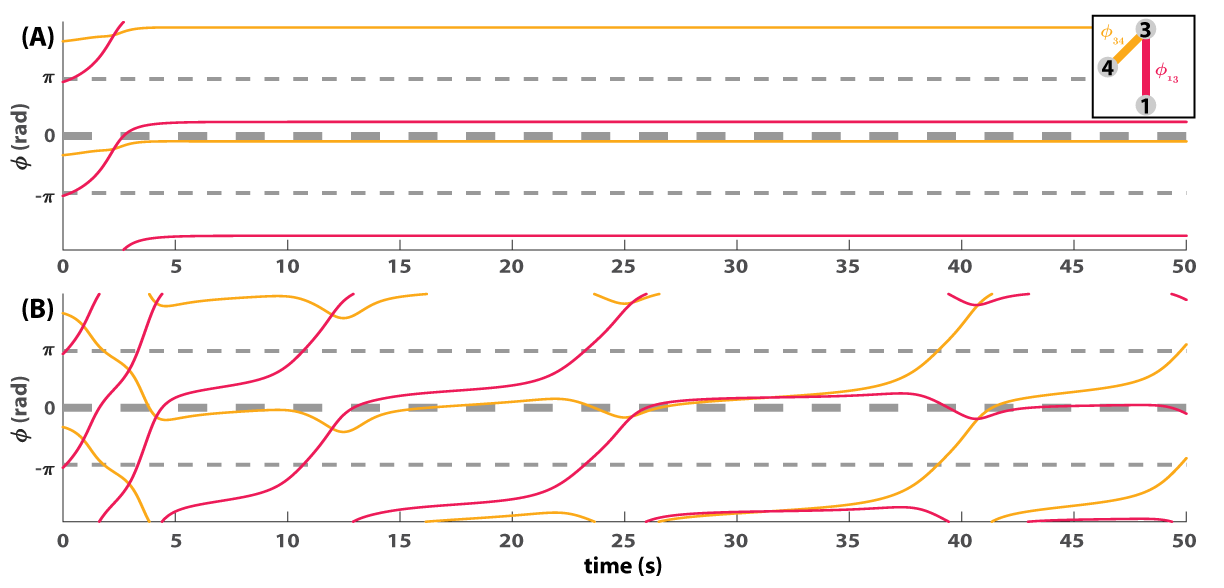}
		\caption{Simulated triadic coordination with (A) $ a_1=a_3=a_4=0.4033 $ and (B) varying natural frequency $ \omega_3 $. }\label{fig:triad_SI}
	\end{fullwidth}
\end{figure}

Here we provide in Figure~\ref{fig:triad_SI} two additional variations of the simulated triadic dynamics shown in Figure~\ref{fig:triad}B. Figure~\ref{fig:triad_SI}A shows what happens when all three oscillators have the identical coupling style, i.e. $ a_1=a_3=a_4 $ and $ b_1=b_3=b_4 $ (keeping the same mean coupling strength as Figure~\ref{fig:triad}B and C). With the symmetry completed restored (in contrast to Figure~\ref{fig:triad}C where only the symmetry between agent 3 and 4 is restored), not only the ``bumps" in $ \phi_{34} $ are gone but also the metastability altogether (at least at the observable time scale). This further illustrates the role of symmetry breaking in understanding the single-trial dynamics. 

Figure~\ref{fig:triad_SI}B shows what happens when agent 3's natural frequency is not constant. A main clue suggesting a non-constant natural frequency is the increasing size of ``bumps" in $ \phi_{34} $ observed in the human behavior (see Figure~\ref{fig:triad}A, the bump in yellow line at 15s was smaller than the one at 25s, and even smaller than the one at 37s) which was accompanied by growing length of the dwells in $ \phi_{13} $ (red trajectory in Figure~\ref{fig:triad}A has three periods of flattening, each one longer than the previous one). This could simply mean that agent 3's ``natural frequency" was moving towards agent 1's and away from agent 4's. In the model, the natural frequencies of agent 1 and 4 are 1.57 and 1.45 Hz respectively. We simply let $ \omega_3 $ increase linearly from 1.2 Hz to 1.7 Hz, instead of being constant (i.e. 1.375 Hz for Figure~\ref{fig:triad}BC and \ref{fig:triad_SI}A), over the course of the trial. The resulted dynamics is shown in Figure~\ref{fig:triad_SI}B. We see the dwells of $ \phi_{13} $ (red line flattening around 7, 17 and 32s) are getting longer over time as the bumps in $ \phi_{34} $ (yellow line) grow (the last bump grows out of itself at 37s and leaves inphase). In fact, at the end of the last dwell (around 37s) $ \phi_{13} $ is no longer metastable in the original sense but begins to oscillate around inphase $ \phi=0 $, whereas $ \phi_{34} $ takes its place at that time and becomes metastable (i.e. after 37s yellow line starts wrapping). 

Gradually increasing natural frequency of agent 3 ($ \omega_3 $) creates two subtle effects in addition to the increasing bump size. The first has already been hinted at that a gradual change of parameter can cause $ \phi_{34} $ to suddenly leave inphase ($ \sim $ 37s yellow line in Figure~\ref{fig:triad_SI}B). In the human trial (Figure~\ref{fig:triad}A), $ \phi_{34} $ had also, after the third bump, left inphase (37s). The difference is that the humans left for antiphase, instead of becoming metastable as for our simple model assuming linearly increasing natural frequency. This suggests that there was, unsurprisingly, more interesting adaptation going on in human movement frequency than just a linear ramping. Another subtle effect is of the same flavor but is concerned with what happens before $ \phi_{34} $ began to dwell at inphase. In the human trial, $ \phi_{34} $ decreased for almost one cycle before it stopped at inphase (0-10s yellow line in Figure~\ref{fig:triad}A). This is not the case with constant frequency (Figure~\ref{fig:triad}B, yellow line, $ \phi_{34} $ immediately increases to inphase after the beginning of the trial), but it is the case with varying frequency (Figure~\ref{fig:triad_SI}B, yellow line, 0-5s). All these show by a very simple example how gradual adaptation in natural frequency may cause sudden changes in coordination patterns.

\subsection{A note on metastability}\label{section:meta}
For intuition, let us assume that there are $ N $ oscillators in a stationary organization defined by $ N-1 $ relative phases, each of which remains near inphase, near antiphase, or wrapping, giving us $ S=3^{N-1} $ different stationary patterns for our model ($ S=2^{N-1} $ for the Kuramoto model because of the lack of antiphase). Now if we look at patterns as sequences of metastable dwells, we could have $ M=\sum_{l=1}^{S} \frac{S!}{(S-l)!} $ patterns of various period $ l $ (with non-repeated spatial configurations in sequence). These of course are not all necessarily reachable by a system, which in itself is an interesting theoretical problem, but still the repertoire $ M $ is much greater than $ S $. This thought experiment shows how metastability contributes to biological complexity in a very significant way.

\subsection{Design of the human experiment}\label{section:expdesign}
The human experiment \cite{Zhang2018} was participated by a total of 120 subjects in 15 ensembles. Each ensemble completed 18 trials (6 trials for each condition $ \delta f=0 $, $ 0.3 $ and $ 0.6 $ Hz) of interaction in a complete network, except one ensemble for which only 7 trials (2 for $ \delta f=0 $ Hz, 2 for $ \delta f=0.3 $ Hz, and 3 for $ \delta f=0.6 $ Hz) were completed due to equipment malfunction. This yields 86 trials for $ \delta f=0 $ Hz, 86 trials for $ \delta f=0.3 $ Hz, and 87 trials for $ \delta f=0.6 $ Hz. See \cite{Zhang2018} for additional details.

\end{document}